\documentclass{article}

\usepackage[scr=boondoxo,scrscaled=1.05]{mathalfa}
\usepackage[pdftex]{graphicx}
\usepackage{amsmath,amssymb,hhline,tikz,mathtools}
\usepackage{enumerate}
\usetikzlibrary{shapes.geometric,decorations.markings}
\usetikzlibrary{calc}
\usepackage[title]{appendix}
\usepackage{etoolbox}
\patchcmd{\appendices}{\quad}{: }{}{}
\usepackage{algorithm}
\usepackage{algorithmicx}
\usepackage{algpseudocode}
\usepackage{rotating}
\usepackage{afterpage}
\usepackage{xstring}
\usepackage{xcolor}
\usepackage{cmap}
\usepackage{mathdots}
\usepackage{colortbl}
\usepackage{booktabs}
\usepackage{textcomp}     
\usetikzlibrary{backgrounds}
\usepackage{pgfplots}
\pgfplotsset{compat=1.12}
\pgfmathsetseed{5112}
\usepackage{multirow}
\usepackage{pgfplotstable}
\usepackage{adjustbox}
\usepackage{diagbox}

\PassOptionsToPackage{hyphens}{url}
\usepackage{hyperref}
\hypersetup{
    pdftitle={Word Embedding Techniques for Malware Classification},
    pdfauthor={Mark Stamp},
    bookmarksnumbered=true,     
    bookmarksopen=true,         
    bookmarksopenlevel=3,       
    colorlinks=true,linkcolor=black,citecolor=black,urlcolor=blue,filecolor=blue,
    pdfstartview=Fit,           
    pdfpagemode=UseOutlines,
    pdfpagelayout=SinglePage
}

\definecolor{darkgreen}{rgb}{0.125,0.5,0.169}

\algrenewcommand{\algorithmiccomment}[1]{{\color{red}{\tt //}\ #1}}
\algnewcommand{\Initialize}[1]{%
  \State \textbf{Initialize:}
  \Statex \hspace*{\algorithmicindent}\parbox[t]{.8\linewidth}{\raggedright #1}
}
\algnewcommand{\Given}[1]{%
  \State \textbf{Given:}
  \Statex \hspace*{\algorithmicindent}\parbox[t]{.8\linewidth}{\raggedright #1}
}

\long\def\symbolfootnotetext[#1]#2{\begingroup%
\def\thefootnote{\fnsymbol{footnote}}\footnotetext[#1]{#2}\endgroup}

\let\oldsqrt\sqrt
\def\sqrt{\mathpalette\DHLhksqrt}
\def\DHLhksqrt#1#2{%
\setbox0=\hbox{$#1\oldsqrt{#2\,}$}\dimen0=\ht0
\advance\dimen0-0.2\ht0
\setbox2=\hbox{\vrule height\ht0 depth -\dimen0}%
{\box0\lower0.4pt\box2}}

\def\clap#1{\hbox to 0pt{\hss#1\hss}}

\allowdisplaybreaks

\def\figureFastSpeed{s}\def\figureSpeed{f}
\let\figureFastSpeed=\figureSpeed
\def\selectFigureSpeed#1#2{
\if\figureSpeed\figureFastSpeed #1\else #2\fi}

\def\srowvecc#1#2{(\!\begin{array}{cc} 
      \noexpandarg\IfBeginWith{#1}{-}{\! #1}{#1}
    & #2\kern-0.5pt\end{array}\!)}
\def\rowvecc#1#2{\left(\!\begin{array}{cc} 
      \noexpandarg\IfBeginWith{#1}{-}{\! #1}{#1}
    & #2\kern-0.5pt\end{array}\!\right)}
\def\rowveccc#1#2#3{\left(\!\begin{array}{ccc} 
      \noexpandarg\IfBeginWith{#1}{-}{\! #1}{#1}
    & #2 
    & #3\kern-0.5pt\end{array}\!\right)}
\def\rowvecccc#1#2#3#4{\left(\!\begin{array}{cccc}
      \noexpandarg\IfBeginWith{#1}{-}{\! #1}{#1}
    & #2 
    & #3 
    & #4\kern-0.5pt\end{array}\!\right)}
\def\srowvecccc#1#2#3#4{\bigl(\!\begin{array}{cccc}
      \noexpandarg\IfBeginWith{#1}{-}{\! #1}{#1}
    & #2 
    & #3 
    & #4\kern-0.5pt\end{array}\!\bigr)}
\def\rowveccccc#1#2#3#4#5{\left(\!\begin{array}{ccccc} 
      \noexpandarg\IfBeginWith{#1}{-}{\! #1}{#1}
    & #2
    & #3
    & #4
    & #5\kern-0.5pt\end{array}\!\right)}
\def\srowvecccccc#1#2#3#4#5#6{(\!\begin{array}{cccccc} 
      \noexpandarg\IfBeginWith{#1}{-}{\! #1}{#1}
    & #2
    & #3
    & #4
    & #5
    & #6\kern-0.5pt\end{array}\!)}
\def\rowvecccccc#1#2#3#4#5#6{\left(\!\begin{array}{cccccc} 
      \noexpandarg\IfBeginWith{#1}{-}{\! #1}{#1}
    & #2
    & #3
    & #4
    & #5
    & #6\kern-0.5pt\end{array}\!\right)}

\def\figureType{*}\def\figureSlowType{slowType}
\def\selectFigureType#1#2{
\if\figureType\figureSlowType #1\else #2\fi}
\makeatletter
\newcommand{\lowsub}[1]{\mathpalette{\raisem@th{#1}}}
\newcommand{\raisem@th}[3]{\raisebox{-#1}{$#2#3$}}
\makeatother
\def\halfthin{\kern 0.083em}

\DeclareMathOperator{\thth}{th}

\def\eref#1{{\color{black}(\ref{#1})}}

\pgfkeys{
    /pgf/number format/precision=2, 
    /pgf/number format/fixed zerofill=true }
    
\pgfplotstableset{
    /color cells/min/.initial=0,
    /color cells/max/.initial=1000,
    /color cells/textcolor/.initial=,
    color cells/.code={%
        \pgfqkeys{/color cells}{#1}%
        \pgfkeysalso{%
            postproc cell content/.code={%
                \begingroup
                \pgfkeysgetvalue{/pgfplots/table/@preprocessed cell content}\value
\ifx\value\empty
\endgroup
\else
                \pgfmathfloatparsenumber{\value}%
                \pgfmathfloattofixed{\pgfmathresult}%
                \let\value=\pgfmathresult
                \pgfplotscolormapaccess
                    [\pgfkeysvalueof{/color cells/min}:\pgfkeysvalueof{/color cells/max}]%
                    {\value}%
                    {\pgfkeysvalueof{/pgfplots/colormap name}}%
                \pgfkeysgetvalue{/pgfplots/table/@cell content}\typesetvalue
                \pgfkeysgetvalue{/color cells/textcolor}\textcolorvalue
                \toks0=\expandafter{\typesetvalue}%
                \xdef\temp{%
                    \noexpand\pgfkeysalso{%
                        @cell content={%
                            \noexpand\cellcolor[rgb]{\pgfmathresult}%
                            \noexpand\definecolor{mapped color}{rgb}{\pgfmathresult}%
                            \ifx\textcolorvalue\empty
                            \else
                                \noexpand\color{\textcolorvalue}%
                            \fi
                            \the\toks0 %
                        }%
                    }%
                }%
                \endgroup
                \temp
\fi
            }%
        }%
    }
}

\makeatletter
\newcommand*\bigcdot{\mathpalette\bigcdot@{.5}}
\newcommand*\bigcdot@[2]{\mathbin{\vcenter{\hbox{\scalebox{#2}{$\m@th#1\bullet$}}}}}
\makeatother

\def\O{{\cal O}}

\def\k{\kern 2.75pt}

\newlength{\xxxxx}
\def\log{\mbox{log}}

\def\z{\phantom{0}}

\usepackage{epsfig}
\usepackage{multicol}
\usepackage{pslatex}
\usepackage{apalike}

\advance\oddsidemargin by -0.45in
\advance\textwidth by 0.9in
\advance\topmargin by -0.6in
\advance\textheight by 1.2in

\begin{document}

\title{Malware Classification with GMM-HMM Models}

\author{Jing Zhao\footnote{jing.zhao@sjsu.edu}
\and
Samanvitha Basole\footnote{s97basole@gmail.com}
\and 
Mark Stamp\footnote{mark.stamp@sjsu.edu}
}

\maketitle

\abstract{Discrete hidden Markov models (HMM) are often applied to malware detection 
and classification problems. However, the continuous analog of discrete HMMs, that is, 
Gaussian mixture model-HMMs (GMM-HMM), are rarely considered in the field of cybersecurity. 
In this paper, we use GMM-HMMs for malware classification and we compare 
our results to those obtained using discrete HMMs. As features, we consider opcode sequences 
and entropy-based sequences. For our opcode features, GMM-HMMs produce results that 
are comparable to those obtained using discrete HMMs, whereas for our entropy-based features, 
GMM-HMMs generally improve significantly on the classification results that we have 
achieved with discrete HMMs.}


\section{Introduction}

Due to COVID-19, businesses and schools have moved their work online and 
some consider the possibility of going online permanently. 
This trend makes cybersecurity more important than ever before.

Malicious software, or malware, is designed to steal private information, 
delete sensitive data without consent, or otherwise disrupt computer systems. 
The study of malware has been active for decades~\cite{malwarehistory}. 
Malware detection and classification are fundamental research topics in malware. 
Traditionally, signature detection has been the most prevalent method for detecting 
malware, but recently, machine learning techniques have proven
their worth, especially for dealing with advanced types of malware. 
Many machine learning approaches have been applied to the malware problem, 
including hidden Markov models (HMM)~\cite{stamp}, $k$-nearest neighbors (KNN)~\cite{knn}, 
support vector machines (SVM)~\cite{svm}, and a wide variety of neural networking
and deep learning techniques~\cite{nnmalware}. 

Each machine learning technique has its own advantages and disadvantages. It is not
the case that one technique is best for all circumstances, since there 
are many different types of malware and many different features that
can be considered. Thus, it is useful to explore different techniques and algorithms in 
an effort to extend our knowledge base for effectively dealing with malware. In this paper, 
we focus on Gaussian mixture model-hidden Markov models (GMM-HMMs), which can
be viewed as the continuous analog of the ever-popular discrete HMM.

Discrete HMMs are well known for their ability to learn important statistical properties from
a sequence of observations. For a sequence of discrete observations, such as the letters that 
comprise a selection of English text, we can train a discrete HMM to determine the parameters
of the (discrete) probability distributions that underlie the training data.
However, some observation sequences are inherently continuous, such as signals
extracted from speech. In such cases,
a discrete HMM is not the ideal tool. 
While we can discretize a continuous signal,
there will be some loss of information. As an alternative to discretization, we can 
attempt to model the continuous probability density functions that underlie 
continuous training data.

Gaussian mixture models (GMM) are probability density functions that are represented 
by weighted sums of Gaussian distributions~\cite{reynolds2009gaussian}. 
By varying the number of Gaussian components and the weight assigned to each, 
GMMs can effectively model a wide variety of continuous probability distributions.
It is possible to train HMMs to learn the parameters of GMMs, and the resulting
GMM-HMM models are frequently used in speech recognition~\cite{rabiner,bansal2008improved},
among many other applications. 

In the field of cybersecurity, GMMs have been used, for example, as a clustering method for 
malware classification~\cite{interrantegaussian}. However, to the best of our knowledge, 
GMM-HMMs are not frequently considered in the context of malware detection or 
classification. In this paper, we apply GMM-HMMs to the malware classification problem, 
and we compare our results to discrete HMMs. Our results indicate that GMM-HMMs
applied to continuous data can yield strong results in the malware domain.

The remainder of this paper is organized as follows. In Chapter~\ref{chap:relatedwork}, 
we discuss relevant related work. Chapter~\ref{chap:foundation} provides background on the 
various models considered, namely, GMMs, HMMs, and GMM-HMMs,
with the emphasis on the latter. Malware classification 
experiments and results based on discrete features
are discussed in Chapter~\ref{chap:experiments}. 
Since GMM-HMMs are more suitable for continuous observations, 
in Chapter~\ref{chap:experiments} we also present a set of malware 
classification experiments based on continuous entropy 
features. We conclude the paper and provide possible directions for future work
in Chapter~\ref{chap:result}.

\section{Related Work}\label{chap:relatedwork}

A Gaussian mixture model (GMM) is a probability density model~\cite{mclachlan2004finite} 
consisting of a weighted sum of multiple Gaussian distributions. The advantage of a Gaussian 
mixture is that it can accurately model a variety of probability distributions~\cite{GAO2020107815}. 
That is, a GMM enables us to model a much more general distribution, as compared to a single 
Gaussian. Although the underlying distribution may not be similar to a Guassian, 
the combination of several Gaussians yields a robust model~\cite{ALFAKIH2020101218}. 
However, the more Gaussians that comprise a model, the costly
the calculation involving the model.

One example of the use of GMMs is distribution estimation of wave elevation in 
the field of oceanography~\cite{GAO2020107815}. GMMs have also been used in the fields of 
anomaly detection~\cite{CHEN20199}, and 
signal mapping~\cite{RAITOHARJU2020107330}. 
As another example, in~\cite{QIAO2019104628}, a GMM is 
used as a classification method to segment brain lesions.
In addition to distribution estimation, GMMs form the basis for a clustering 
method in~\cite{gallopj}. 

As the name suggests, a discrete hidden Markov model (HMM) includes a ``hidden''
Markov process and a series of observations that are probabilistically related
to the hidden states. An HMM can be trained based on an observation sequence,
and the resulting model can be used to score other observation sequences.
HMMs have found widespread use in signal processing, and HMMs are particularly
popular in the area of speech recognition~\cite{1333078}. Due to their robustness 
and the efficiency, HMMs are also widely used in medical areas, such 
as sepsis detection~\cite{6680664} and human brain studies based on 
functional magnetic resonance imaging~\cite{DANG201787}. Motion recognition 
is another area where HMMs play a vital role; specific examples include
recognizing dancing moves~\cite{hmmdancing} and 3D gestures~\cite{hmm3dgestures}.

Gaussian mixture model-HMMs (GMM-HMM)
are also widely used in classification problems. Given the flexibility 
of GMMs, GMM-HMMs are popular for dealing with complex patterns 
underlying sequences of observations. For example, Yao et al.~\cite{YAO2020102711}
use GMM-HMMs to classify network traffic from different protocols. GMM-HMMs 
have also been used in motion detection---for complex poses, GMM-HMMs 
outperform discrete HMMs~\cite{ZHANG2020106603}.


\section{Background\label{chap:foundation}}

In this section, we first introduce the learning techniques used in this paper---specifically, 
we discuss Gaussian mixture models, HMMs, and GMM-HMMs. We then discuss GMM-HMMs
in somewhat more detail, including various training and parameter selection issues,
and we provide an illustrative example of GMM-HMM training.

\subsection{Gaussian Mixture Models}

As mentioned above, a GMM is a probabilistic model that combines 
multiple Gaussian distributions. Mathematically, the probability density function 
of a GMM is a weighted sum of~$M$ Gaussian probability density functions. 
The formulation of a GMM can be written as~\cite{gmm} 
$$
    P(x\,|\,\lambda) = \sum_{x=i}^{M}\omega_i\, g(x\,|\,\mu_i, \Sigma_i),
$$
where~$x$ is a~$D$-dimensional vector and~$\omega_i$ is the weight assigned to 
the $i^{\thth}$ Gaussian component, with the mixture weights summing to one.
Here, $\mu_i$ and~$\Sigma_i$ are the mean and the covariance matrix of 
the~$i^{\thth}$ component of the GMM, respectively. Each component of a
GMM is a multivariate Gaussian distribution of the form
$$
    g(x\,|\,\mu_i, \Sigma_i)
      = \frac{1}{(2\pi)^{\frac{D}{2}}\,|\,\Sigma_i\,|^{\frac{1}{2}}}\,
      	e^{-\frac{1}{2}(x-\mu_i)'\Sigma_{i}^{-1}(x-\mu_i)} .
$$

\subsection{Discrete HMM}\label{sec:hmm}

In this paper, we use the notation in Table~\ref{tab:HMMnotation}
to describe a discrete HMM.
This notation is essentially the same as that given in~\cite{stamp}. 
An HMM, which we denote as~$\lambda$, 
is defined by the matrices~$A$, $B$, and~$\pi$,
and hence we have~$\lambda = (A, B, \pi)$.

\begin{table}[!htb]
  \centering
  \caption{Discrete HMM notation}\label{tab:HMMnotation}
  \resizebox{0.475\textwidth}{!}{
  \begin{tabular}{cl} \midrule\midrule
    \textbf{Notation} & \hspace*{0.5in}\textbf{Explanation}\\ \midrule
    $T$ & Length of the observation sequence\\
    $\O$ & Observation sequence, $\O_0,\O_1,\ldots,\O_{T-1}$ \\
    $N$ & Number of states in the model\\
    $K$ & Number of distinct observation symbols\\
    $Q$ & Distinct states of the Markov process, $q_0,q_1,\ldots,q_{N-1}$\\
    $V$ & Observable symbols, assumed to be $0,1,\ldots,K-1$\\
    $\pi$ & Initial state distribution, $1 \times N$\\
    $A$ & State transition probabilities, $N \times N$\\
    $B$ & Observation probability matrix, $N \times K$\\
    \midrule\midrule
  \end{tabular}
  }
\end{table}

We denote the elements in row~$i$ and column~$j$ 
of~$A$ as~$a_{ij}$.
The element~$a_{ij}$ of the~$A$ matrix is given by
$$
  a_{ij} = P(\mbox{state }q_j \mbox{ at } t+1\,|\, \mbox{ state } q_i \mbox{ at } t) .
$$

The~$(i,j)$ element of~$B$ is denoted in a slightly unusual form as~$b_i(j)$.
In a discrete HMM, 
row~$i$ of~$B$ represents the (discrete) probability distribution 
of the observation symbols when underlying Markov process is in
(hidden) state~$i$. Specifically, each element of~$B=\{b_i(j)\}$ matrix is given 
by
$$
  b_i(j)=P(\mbox{observation } j \mbox{ at } t\, |\, \mbox{ state } q_i \mbox{ at } t) .
$$

The HMM formulation can be used to is solve the following three problems~\cite{stamp}.
\begin{enumerate}
\item Given an observation sequence~$\O$ and a 
model~$\lambda$ of the form~$\lambda = (\pi, A, B)$, calculate the probability of the observation sequence.
That is, we can score an observation sequence against a given model.
\item Given a model~$\lambda = (\pi, A, B)$ and an observation sequence~$\O$, 
find the ``best'' state sequence, where best is defined to be the sequence
that maximizes the expected number of correct states. That is, we can uncover
the hidden state sequence.
\item Given an observation sequence~$\O$, 
determine a model~$\lambda = (A, B, \pi)$ that maximizes~$P(\O\,|\,\lambda)$. 
That is, we can train a model for a given observation sequence.
\end{enumerate}

In this research, we are interested in problems~1 and~3. Specifically, we train models,
then we test the resulting models by scoring observation sequences. The solution to
problem~2 is of interest in various NLP applications, for example. For the sake
of brevity, we omit the details of training and scoring with discrete HMMs;
see~\cite{stamp} or~\cite{rabiner} for more information.

\subsection{GMM-HMM}

The structure of a GMM-HMM is similar to that of a discrete HMM. 
However, in a GMM-HMM, the~$B$ matrix is much different, since
we are dealing with a mixture of (continuous) Gaussian distributions, rather
than the discrete probability distributions a discrete HMM. In a GMM-HMM, 
the probability of an observation at a given state is determined by a probability 
density function that is defined by a GMM. Specifically, the probability density 
function of observation~$\O_t$ when the model is in state~$i$ is given by
\begin{equation}\label{eq:g}
    P_i(\O_t) = \sum_{m=1}^{M}c_{im}g(\O_t\,|\,\mu_{im}, \Sigma_{im}) ,
\end{equation}
for~$i\in\{1, 2, \ldots, N\}$ and~$t\in\{0, 1, \ldots, T-1$, where
$$
    \sum_{m=1}^{M}c_{im} = 1 \mbox{ for } i\in\{1, 2, \ldots, N\} .
$$
Here, $M$ is the number of Gaussian mixtures components, 
$c_{im}$ is the mixture coefficient or the weight of $m^{\thth}$ 
Gaussian mixture at state $i$, while~$\mu_{im}$ and~$\Sigma_{im}$ 
are the mean vector and covariance matrix for the $m^{\thth}$ 
Gaussian mixture at state $i$. 
We can rewrite~$g$ in equation~\eref{eq:g} as
$$
   g(\O_t\,|\,\mu_{im}, \Sigma_{im})
      = \frac{1}{(2\pi)^{\frac{D}{2}}\,|\,\Sigma_{im}\,|^{\frac{1}{2}}}\,
    	e^{-\frac{1}{2}(\O_t-\mu_{im})'\Sigma_{im}^{-1}(\O_t-\mu_{im})},
$$
where~$D$ is the dimension of each observation. 
In a GMM-HMM, the~$A$ and~$\pi$ matrices are the same as in a discrete HMM. 

The notation for a GMM-HMM is given in Table~\ref{tab:GMM-HMMnotation}.
This is inherently more complex than a discrete HMM,
due to the presence of the~$M$ Gaussian distributions.
Note that a GMM-HMM is defined by the 5-tuple
$$
  \lambda = (A, \pi, c, \mu, \Sigma) .
$$

\begin{table}[!htb]
  \caption{GMM-HMM notation}\label{tab:GMM-HMMnotation}
  \centering
  \resizebox{0.475\textwidth}{!}{
  \begin{tabular}{cl} \midrule\midrule
    \textbf{Notation} & \hspace*{0.5in}\textbf{Explanation}\\ \midrule
    $T$ & Length of the observation sequence\\
    $\O$ & Observation sequence, $\O_0,\O_1,\ldots ,\O_{T-1}$ \\
    $N$ & Number of states in the model\\
    $M$ & Number of Gaussian components\\
    $D$ & Dimension of each observation\\
    $\pi$ & Initial state distribution, $1 \times N$\\
    $A$ & State transition matrix, $N \times N$\\
    $c$ & Gaussian mixture weight at each state, $N \times M$\\
    $\mu$ & Means of Gaussians at each state, $N \times M \times D$\\
    $\Sigma$ & Covariance of Gaussian mixtures, $N \times M \times D \times D$\\ 
    \midrule\midrule
  \end{tabular}
  }
\end{table}

Analogous to a discrete HMM, we can solve the same three problems with
a GMM-HMM. However, the process
used for training and scoring with a GMM-HMM differ significantly
as compared to a discrete HMM.


\subsection{GMM-HMM Training and Scoring}

To use a GMM-HMM to classify malware samples, 
we need to train a model, then use the resulting 
model to score samples---see the discussion of problems~1
and~3 in Section~\ref{sec:hmm}, above. 
In this section, we discuss scoring and training in the context
of a GMM-HMM in some detail. We begin with the simpler
problem, which is scoring.

\subsubsection{GMM-HMM Scoring}

Given a GMM-HMM, which is defined by the 
5-tuple of matrices~$\lambda = (A, \pi, c, \mu, \Sigma)$,
and a sequence of observations~$\O=\{\O_0, \O_1, \ldots, \O_{T-1}\}$, 
we want to determine~$P(\O\,|\,\lambda)$. The forward algorithm, 
which is also known as the~$\alpha$-pass, 
can be used to efficiently compute~$P(\O\,|\,\lambda)$.

Analogous to a discrete HMM as discussed in~\cite{stamp}, 
in the $\alpha$-pass of a GMM-HMM, we 
define
$$
  \alpha_t(i) = P(\O_0, \O_1,\ldots,\O_t, x_t=q_i\,|\,\lambda) ,
$$ 
that is, $\alpha_t(i)$ is the probability of the partial sequence of 
observation up to time~$t$, ending in state~$q_i$ at time~$t$. 
The desired probability is given by
$$
    P(\O\,|\,\lambda) = \sum\limits_{i=0}^{N-1} \alpha_{T-1}(i) .
$$
The~$\alpha_t(i)$ can be computed recursively as
\begin{equation}\label{eq:alpha}
    \alpha_t(i) = \biggl(\sum\limits_{j=0}^{N-1}\alpha_{t-1}(i)a_{ji}\biggr)b_i(\O_t) .
\end{equation}
At time~$t=0$, from the definition it is clear that we have~$\alpha_0(i) = \pi_ib_i(\O_0)$.

In a discrete HMM, $b_i(\O_t)$ gives the probability of observing~$\O_t$ 
at time~$t$ when the underlying Markov process is in state~$i$. 
In a GMM-HMM, however, simply replacing~$b_i(\O_t)$ in~\eref{eq:alpha}
by the GMM pdf corresponds to a point value of a continuous distribution. 
To obtain the desired probability, as discussed in~\cite{hmmcon},
we must integrate over of a small region 
around observation~$\O_t$, that is, we compute
\begin{equation}\label{eqa:continous-probability}
    b_i(\O_t) = \int_{\O_t-\epsilon}^{\O_t+\epsilon} p_i(\O_t\,|\,\theta_i)\,d\O,
\end{equation}
where~$\theta_i$ consists of the parameters~$c_i$, $\mu_i$ and~$\Sigma_i$
of the GMM, and~$\epsilon$ is a (small) range parameter. 

\subsubsection{GMM-HMM Training}

The forward algorithm or~$\alpha$-pass calculates the probability of observing the sequence 
from the beginning up to time~$t$. There is an analogous backwards pass
or~$\beta$-pass that calculates the probability of the tail
of the sequence, that is, the sequence from~$t+1$ to the end. 
In the~$\beta$-pass, we define
$$
  \beta_t(i) = P(\O_{t+1}, \O_{t+2}, \ldots, \O_{T-1}\,|\,x_i = q_i, \lambda) .
$$ 
The~$\beta_t(i)$ can be compute recursively via
$$
    \beta_t(i) = \sum\limits_{j=0}^{N-1}a_{ij}b_j(\O_t)\beta_{t+1}(j)
$$
where we the initialization is~$\beta_{T-1}(i) = 1$, which follows from the definition.

In a discrete HMM, to re-estimate the state transitions in the~$A$
matrix, we first define
$$
  \gamma_t(i, j) = P(x_t=q_i, x_{t+1}=q_j\,|\,\O, \lambda)
$$ 
which is the probability of being in state~$q_i$ at time~$t$ and 
transiting to state~$q_j$ at time~$t+1$. Using the~$\alpha$-pass and 
the~$\beta$-pass, we can efficiently compute~$\gamma_t(i, j)$;
see~\cite{stamp} for the details. The sum of these ``di-gamma'' values 
with respect to the transiting states gives the probability of the observation 
being in state $q_i$ at time $t$, which we define as~$\gamma_t(i)$.
That is,
$$
  \gamma_t(i) = \sum\limits_{j=1}^{N}\gamma_t(i, j) .
$$ 
Thus, we can re-estimate the elements of the~$A$ matrix in a discrete HMM as
$$
  a_{ij} = \frac{\sum\limits_{t=0}^{T-2}\gamma_t(i, j)}{\sum\limits_{t=0}^{T-2}\gamma_t(i)}
$$

To train a GMM-HMM, we use an analogous strategy as that 
used for the discrete HMM. The GMM-HMM analog of the di-gamma form is 
$$
    \gamma_t(j, k) = P(x_t = q_j\,|\, k, \O, \lambda),
$$
where~$t = 0,1,\ldots,T-2$, and~$j=1,2,\ldots,N$, and 
we have~$k=1,2,\ldots,M$. 
Here, $\gamma_t(j, k)$ represents the probability of being state~$q_j$ at 
time~$t$ with respect to the~$k^{\thth}$ Gaussian mixture. 
According to~\cite{rabiner}, these~$\gamma_t(j, k)$ are computed as
$$
    \gamma_{t}(j, k) = 
    \frac{\alpha_t(j)\beta_t(j)}{\sum\limits_{j=1}^{N} \alpha_{t}(j)\beta_{t}(j)}
    \cdot
    \frac{c_{jk}N(\O_t\,|\,\mu_{jk},\Sigma_{jk})}{
    	\sum\limits_{m=1}^{M}c_{jm}N(\O_t\,|\,\mu_{jm}, \Sigma_{jm})}
$$
where the~$\alpha_t(j)$ and~$\beta_t(j)$ are defined above, and~$c_{jk}$ is the weight 
of the $k^{\thth}$ Gaussian mixture component.

The re-estimates for the weights~$c_{jk}$ of the Gaussian mixtures are given by
\begin{equation}\label{eq:cjk}
    \hat{c}_{jk} 
    	= \frac{\sum\limits_{t=0}^{T-1}\gamma_{t}(j, k)}{
		    \sum\limits_{t=0}^{T-1}\sum\limits_{k=1}^{M}\gamma_{t}(j, k)} ,
\end{equation}
for~$j=1,2,\ldots,N$ and~$k=1,2,\ldots,M$; see~\cite{6769559} and~\cite{rabiner}
for additional details. The numerator in~\eref{eq:cjk}
can be interpreted as the expected number of transitions from
state~$q_j$ as determined by the~$k^{\thth}$ Gaussian mixture 
while the denominator can viewed as the expected transitions from 
state~$q_j$ given by the~$M$ Gaussian mixtures.
Accordingly, the re-estimation for~$\mu_{jk}$ and~$\Sigma_{jk}$ are of the form 
$$
    \hat{\mu}_{\kern-1pt jk} = \frac{\sum\limits_{t=0}^{T-1}
    	\gamma_{t}(j, k)\O_t}{\sum\limits_{t=0}^{T-1}\gamma_{t}(j, k)}
$$
and 
$$
    \hat{\Sigma}_{jk} = \frac{\sum\limits_{t=0}^{T-1}
    	\gamma_{t}(j, k)(\O_t - \mu_{jk})(\O_t - \mu_{jk})'}{\sum\limits_{t=0}^{T-1}\gamma_t(j, k)},
$$
for $i=1,2,\ldots,N$ and $k=1,2,\ldots,M$.

\subsection{GMM-HMM Example}

As an example to illustrate a GMM-HMM, we train a model on English text, 
which is a classic example for discrete HMMs~\cite{cave}.
With~$N=2$ hidden states and~$M=27$ observation symbols
(corresponding the the~26 letters and word-space), a discrete HMM trained on 
English text will have one hidden state corresponding to consonants,
while the other hidden state corresponds to vowels. That the model can
make this key distinction is a good example of learning, since a priori
no information is provided regarding the differences between the observations.
We consider this same experiment using a GMM-HMM to see how 
this model compares to a discrete HMM. 

The English training data is from the ``Brown corpus''~\cite{english-text}, 
and we convert all letters to lowercase and remove punctuation, numbers,
and other special symbols, leaving only~26 letters and word-spaces. 
For our GMM-HMM training,
we set~$N = 2$, $M = 6$ (i.e., we have a mixture model 
consisting of~6 Gaussians) 
and~$T = 50000$. The~$A$ matrix 
is~$N\times N$, $\pi$ is~$1\times N$, both of which are row stochastic,
and initialized to approximately uniform. The parameter~$c$ represents the weights of 
the mixture components and is initialized with row stochastic values, also
approximately uniform. We use the global mean value (i.e., the mean of all observations) 
and global variance to initialize~$\mu$ and~$\Sigma$. Note that 
each Gaussian is initialized with the same mean and variance.

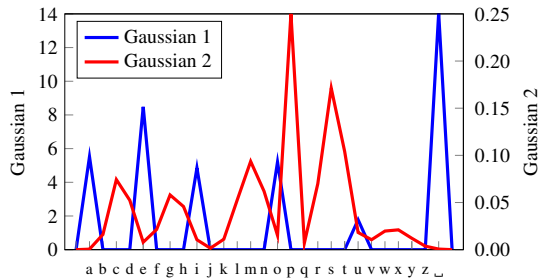
\begin{figure}[!htb]
      \centering
      \begin{tikzpicture}[scale=0.75]
\begin{axis}[width=0.6\textwidth,
		   height=0.4\textwidth,
                    symbolic x coords={B,a,b,c,d,e,f,g,h,i,j,k,l,m,n,o,p,q,r,s,t,u,v,w,x,y,z,S,D},
                    xtick={a,b,c,d,e,f,g,h,i,j,k,l,m,n,o,p,q,r,s,t,u,v,w,x,y,z,S},
                    xticklabels={a,b,c,d,e,f,g,h,i,j,k,l,m,n,o,p,q,r,s,t,u,v,w,x,y,z,\texttt{\char32}},
	 	   x tick label style={
			text height=2ex,
			font=\footnotesize,
    			inner sep=1mm},
		   xtick pos=bottom,
		   ytick pos=left,
	 	   y tick label style={
    		 	/pgf/number format/.cd,
   			fixed,
   			fixed zerofill,
    			precision=0,
			/tikz/.cd},
		   ytick={0,2,4,6,8,10,12,14},
  		   scaled y ticks=false,
		   enlarge x limits=0.03,
                    ymin=0.0,
                    ymax=14.0,
                    ylabel={Gaussian~1}] 
\addplot[color=blue,ultra thick,no marks] coordinates {
(B,0)
(a,5.519725497)
(b,0)
(c,0)
(d,0)
(e,8.48262992)
(f,0)
(g,0)
(h,0)
(i,4.854688716)
(j,0)
(k,0)
(l,0)
(m,0)
(n,0)
(o,5.206356336)
(p,0)
(q,0)
(r,0)
(s,0)
(t,0)
(u,1.777567119)
(v,0)
(w,0)
(x,0)
(y,0)
(z,0)
(S,14.05330035)
(D,0)
}; \label{cv_plot}
\end{axis}
%
%
\begin{axis}[width=0.6\textwidth,
		   height=0.4\textwidth,
		   axis y line*=right,
                    symbolic x coords={B,a,b,c,d,e,f,g,h,i,j,k,l,m,n,o,p,q,r,s,t,u,v,w,x,y,z,S,D},
		   xticklabel=\empty,
		   xtick pos=bottom,
	 	   y tick label style={
    		 	/pgf/number format/.cd,
			1000 sep={},
   			fixed,
   			fixed zerofill,
    			precision=2,
			/tikz/.cd,
			text width=width("$1000$"),align=right},
		   ytick={0.00,0.05,0.10,0.15,0.20,0.25},
		   enlarge x limits=0.03,
                    ymin=0,
                    ymax=0.25,
                    ylabel={Gaussian~2},
                    legend pos = north west] 
\addlegendimage{/pgfplots/refstyle=cv_plot,ultra thick}\addlegendentry{Gaussian~1}
\addplot[color=red,ultra thick,no marks] coordinates {
(B,0)
(a,0.000535443)
(b,0.016131972)
(c,0.074292007)
(d,0.052366317)
(e,0.007906248)
(f,0.021436616)
(g,0.058004938)
(h,0.04574778)
(i,0.010496098)
(j,0.001557956)
(k,0.01106361)
(l,0.052705526)
(m,0.093576243)
(n,0.061844403)
(o,0.015215668)
(p,0.25)
(q,0.007114447)
(r,0.069962765)
(s,0.171291336)
(t,0.104064337)
(u,0.018234566)
(v,0.010545092)
(w,0.019723427)
(x,0.021024729)
(y,0.012087093)
(z,0.003746638)
(S,0.000626167)
(D,0)
}; \addlegendentry{Gaussian~2}
\end{axis}
\end{tikzpicture}
      \caption{English letter distributions in each state}\label{fig:letDist}
\end{figure}

We train~100 of these GMM-HMM models, each with different random initializations. 
As the observations are discrete symbols, the probability of each observation in 
state~$i$ at time~$t$ is estimated by the probability density function. 
The best of the trained models clearly shows that the GMM-HMM 
technique is able to 
successfully group the vowels into one state. This can be seen from 
Figure~\ref{fig:letDist}. Note that in Figure~\ref{fig:letDist},
word-space is represented by the symbol~``\texttt{\char32}''.

Figure~\ref{fig:letDist} clearly shows that all vowels (and word space) 
belong to the first state. 
Table~\ref{tab:tab1} lists the mean value for each Gaussian mixture
in the trained model. The mean value of each Gaussian mixture component 
corresponds to the encoded value of each observation symbol. 

\begin{table}
\centering
\caption{Mean of each Gaussian mixture in each state}\label{tab:tab1}
\resizebox{0.475\textwidth}{!}{
\begin{tabular}{c|cccccc} 
 \midrule\midrule
 \multirow{2}{*}{\textbf{State}} & \multicolumn{6}{c}{\textbf{Gaussian}} \\
                        & 1 & 2 & 3 & 4 & 5 & 6 \\ 
 \midrule
 0 & 26.00 & 14.00 & \z8.00 & \z4.00 & 20.00 & \z0.00\\
 1 & 22.60 &  \z6.31 & 15.00 & 12.08 & \z2.31 & 18.14\\ 
 \midrule\midrule
\end{tabular}
}
\end{table}

In this example, since we know the number of vowels beforehand, we have set the 
number of Gaussian mixture components to~6 (i.e., 5 vowels and word-space). 
In practice, we generally do not know the true number of hidden states, 
in which case we would need to experiment with different numbers of Gaussians.
In general, machine learning and deep learning requires a significant degree of 
experimentation, so it is not surprising that we might need to
fine tune our models.

\section{Malware Experiments}\label{chap:experiments}

In this section, we fist introduce the dataset used in our experiments,
followed by two distinct sets of experiments. In our first set of experiments,
we compare the performance of discrete HMMs and GMM-HMMs
using opcode sequences as our features. In our second set of experiments,
we consider entropy sequences, which serve to illustrate the
strength of the GMM-HMM technique.

\subsection{Dataset}

In all of our experiments, we consider 
three malware families, namely, Winwebsec, Zbot, and Zeroaccess.

\begin{description}
\item[Winwebsec] is a type of Trojan horse in the Windows operating system. 
It attempts to install malicious programs by displaying fake links to 
bait users~\cite{Win32Win35:online}. 
\item[Zbot] is another type of Trojan that 
tries to steal user information by attaching executable files to
spam email messages~\cite{PWSWin3261:online}. 
\item[Zeroaccess] also tries steal information, and 
it can also cause other malicious actions, such as downloading malware or opening a backdoor~\cite{zeroacce66:online}.
\end{description}

Table~\ref{tab:tab2} lists the number of samples of each malware family
in our dataset. These families are part of the Malicia dataset~\cite{malicia-dataset}
and have been used in numerous previous malware studies.

\begin{table}
\centering
\caption{Number of samples in each malware family}\label{tab:tab2}
\begin{tabular}{ c|c } 
 \midrule\midrule
 \textbf{Family} & \textbf{Samples } \\ 
 \midrule
 Winwebsec & 4360 \\
 Zeroaccess & 2136 \\ 
 Zbot& 1305\\ \midrule
 Total & 7801 \\
 \midrule\midrule
\end{tabular}
\end{table}

The samples of each malware family are split into~80\%\ for training and~20\%\ for testing. 
We train models on one malware family, and test the resulting model 
separately against the other two families. Note that each of these
experiments is a binary classification problem.

We use the area under the ROC curve (AUC) as our measure of success.
The AUC can be interpreted as the probability that a randomly selected 
positive sample scores higher than a randomly selected negative sample~\cite{rocBrad}.
We perform 5-fold cross validation, and the average AUC 
from the~5 folds is the numerical result that we use for comparison. 

\subsection{Opcode Features}

For our first set of malware experiments, we compare a 
discrete HMM and GMM-HMM using mnemonic opcode sequences
as features.
To encode the input, we disassemble each executable, then extract
the opcode sequence. We retain the most frequent~30 opcodes 
with all remaining opcodes lumped together into a single ``other'' category,
giving us a total of~31 distinct observations. The percentage of opcodes
that are among the top~30 most frequent are listed in Table~\ref{tab:tab3}.

\begin{table}
\centering
\caption{Percentage of top~30 opcodes}\label{tab:tab3}
\begin{tabular}{ c|c } 
 \midrule\midrule
 \textbf{Family} & \textbf{Top~30 opcodes} \\ 
 \midrule
 Winwebsec & 96.9\% \\
 Zeroaccess & 95.8\% \\ 
 Zbot& 93.4\%\\
 \midrule\midrule
\end{tabular}
\end{table}

For training, we limit the length of the observation sequence to~$T=100000$, and
for the discrete HMM, we let~$N=2$.
For the GMM-HMM, we we experiment with the number of
Gaussian mixtures ranging from~$M=2$ to~$M=5$.  

As mentioned above, we train a model with one malware family and test with 
the other two malware families individually (i.e., in binary classification mode). 
To test each model's performance, we use one hundred samples from 
both families in the binary classification. 

We initialize~$\pi$ and~$A$ to be approximately uniform, as well as making
them row stochastic. For each discrete HMM, the~$B$ matrix is initialize similarly,
while for each GMM-HMM, the mean values and the covariance are initialized 
with the global mean value and the global covariance of all training samples. 

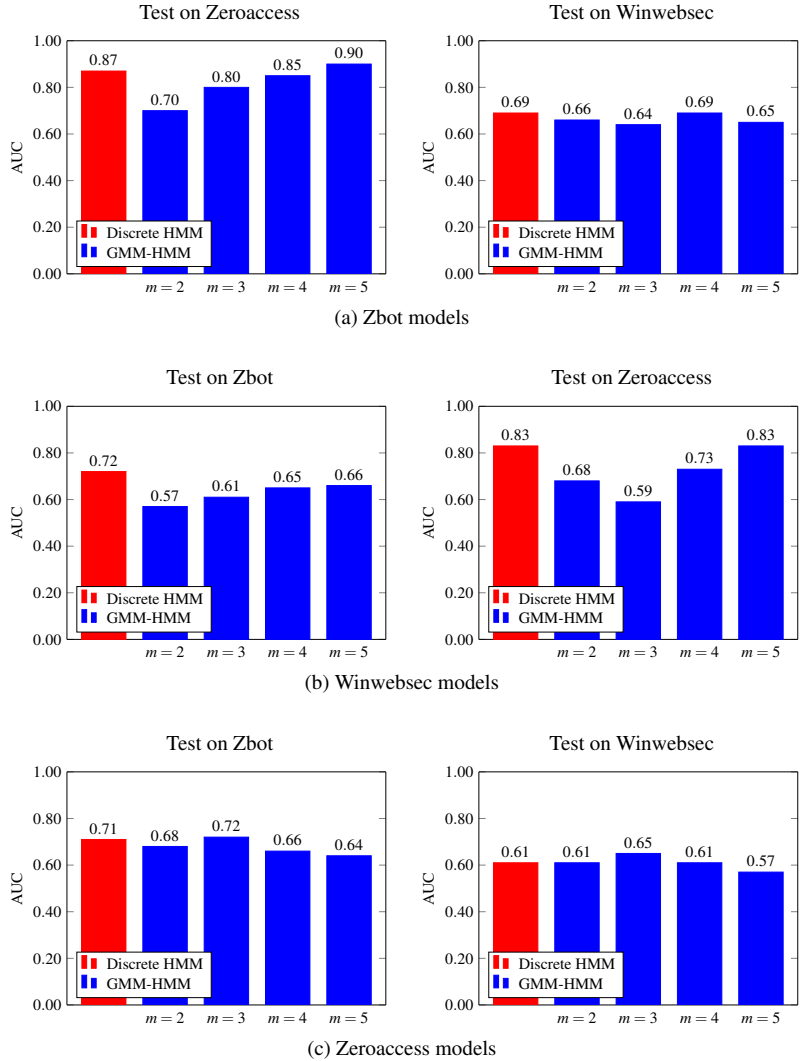
\begin{figure*}[!htb]
      \centering
      \begin{tabular}{cc}
      \hspace*{0.25in}{\footnotesize Test on Zeroaccess}
      & 
      \hspace*{0.25in}{\footnotesize Test on Winwebsec}
      \\ \\[-2.5ex]
      \begin{tikzpicture}[scale=0.6, every node/.style={scale=1.0}]
\begin{axis}[
        width  = 0.6*\textwidth,
        height = 6.75cm,
        ymin=0.0,ymax=1.0,
        ytick={0.0,0.2,0.4,0.6,0.8,1.0},
        separate axis lines,
        major x tick style = transparent,
        ybar=5*\pgflinewidth,
        every axis plot/.append style={
          ybar,
          bar width=28.0pt,
          bar shift=0pt,
          fill
        },
        ylabel = {AUC},
        xtick={1,2,3,4,5},
        x tick style={draw=none},
        xticklabels={
,
$m=2$,
$m=3$,
$m=4$,
$m=5$,
},
	y tick label style={
    		/pgf/number format/.cd,
   		fixed,
   		fixed zerofill,
    		precision=2},
        x tick label style={
		inner sep=0mm},
        nodes near coords,
        every node near coord/.style={text=black},
        every node near coord/.append style={
								   /pgf/number format/.cd,
								   fixed,
								   fixed zerofill,
								   precision=2},
        enlarge x limits=0.15,
        legend cell align=left,
        legend style={
		legend pos = south west,
                	column sep=1ex
        },
]
\addplot[red]coordinates {(1,0.87)};
\addplot[blue]coordinates {(2,0.70)};
\addplot[blue,forget plot]coordinates {(3,0.80)};
\addplot[blue,forget plot]coordinates {(4,0.85)};
\addplot[blue,forget plot]coordinates {(5,0.90)};
\addlegendentry{Discrete HMM}
\addlegendentry{GMM-HMM}
\end{axis}
\end{tikzpicture}
      &
      \begin{tikzpicture}[scale=0.6, every node/.style={scale=1.0}]
\begin{axis}[
        width  = 0.6*\textwidth,
        height = 6.75cm,
        ymin=0.0,ymax=1.0,
        ytick={0.0,0.2,0.4,0.6,0.8,1.0},
        separate axis lines,
        major x tick style = transparent,
        ybar=5*\pgflinewidth,
        every axis plot/.append style={
          ybar,
          bar width=28.0pt,
          bar shift=0pt,
          fill
        },
        ylabel = {AUC},
        xtick={1,2,3,4,5},
        x tick style={draw=none},
        xticklabels={
,
$m=2$,
$m=3$,
$m=4$,
$m=5$,
},
	y tick label style={
    		/pgf/number format/.cd,
   		fixed,
   		fixed zerofill,
    		precision=2},
        x tick label style={
		inner sep=0mm},
        nodes near coords,
        every node near coord/.style={text=black},
        every node near coord/.append style={
								   /pgf/number format/.cd,
								   fixed,
								   fixed zerofill,
								   precision=2},
        enlarge x limits=0.15,
        legend cell align=left,
        legend style={
		legend pos = south west,
                	column sep=1ex
        },
]
\addplot[red]coordinates {(1,0.69)};
\addplot[blue]coordinates {(2,0.66)};
\addplot[blue,forget plot]coordinates {(3,0.64)};
\addplot[blue,forget plot]coordinates {(4,0.69)};
\addplot[blue,forget plot]coordinates {(5,0.65)};
\addlegendentry{Discrete HMM}
\addlegendentry{GMM-HMM}
\end{axis}
\end{tikzpicture} 
      \\
      \multicolumn{2}{c}{\footnotesize (a) Zbot models}
      \\ \\[-0.25ex]
      \hspace*{0.25in}{\footnotesize Test on Zbot}
      & 
      \hspace*{0.25in}{\footnotesize Test on Zeroaccess}
      \\ \\[-2.5ex]
      \begin{tikzpicture}[scale=0.6, every node/.style={scale=1.0}]
\begin{axis}[
        width  = 0.6*\textwidth,
        height = 6.75cm,
        ymin=0.0,ymax=1.0,
        ytick={0.0,0.2,0.4,0.6,0.8,1.0},
        separate axis lines,
        major x tick style = transparent,
        ybar=5*\pgflinewidth,
        every axis plot/.append style={
          ybar,
          bar width=28.0pt,
          bar shift=0pt,
          fill
        },
        ylabel = {AUC},
        xtick={1,2,3,4,5},
        x tick style={draw=none},
        xticklabels={
,
$m=2$,
$m=3$,
$m=4$,
$m=5$,
},
	y tick label style={
    		/pgf/number format/.cd,
   		fixed,
   		fixed zerofill,
    		precision=2},
        x tick label style={
		inner sep=0mm},
        nodes near coords,
        every node near coord/.style={text=black},
        every node near coord/.append style={
								   /pgf/number format/.cd,
								   fixed,
								   fixed zerofill,
								   precision=2},
        enlarge x limits=0.15,
        legend cell align=left,
        legend style={
		legend pos = south west,
                	column sep=1ex
        },
]
\addplot[red]coordinates {(1,0.72)};
\addplot[blue]coordinates {(2,0.57)};
\addplot[blue,forget plot]coordinates {(3,0.61)};
\addplot[blue,forget plot]coordinates {(4,0.65)};
\addplot[blue,forget plot]coordinates {(5,0.66)};
\addlegendentry{Discrete HMM}
\addlegendentry{GMM-HMM}
\end{axis}
\end{tikzpicture}
      &
      \begin{tikzpicture}[scale=0.6, every node/.style={scale=1.0}]
\begin{axis}[
        width  = 0.6*\textwidth,
        height = 6.75cm,
        ymin=0.0,ymax=1.0,
        ytick={0.0,0.2,0.4,0.6,0.8,1.0},
        separate axis lines,
        major x tick style = transparent,
        ybar=5*\pgflinewidth,
        every axis plot/.append style={
          ybar,
          bar width=28.0pt,
          bar shift=0pt,
          fill
        },
        ylabel = {AUC},
        xtick={1,2,3,4,5},
        x tick style={draw=none},
        xticklabels={
,
$m=2$,
$m=3$,
$m=4$,
$m=5$,
},
	y tick label style={
    		/pgf/number format/.cd,
   		fixed,
   		fixed zerofill,
    		precision=2},
        x tick label style={
		inner sep=0mm},
        nodes near coords,
        every node near coord/.style={text=black},
        every node near coord/.append style={
								   /pgf/number format/.cd,
								   fixed,
								   fixed zerofill,
								   precision=2},
        enlarge x limits=0.15,
        legend cell align=left,
        legend style={
		legend pos = south west,
                	column sep=1ex
        },
]
\addplot[red]coordinates {(1,0.83)};
\addplot[blue]coordinates {(2,0.68)};
\addplot[blue,forget plot]coordinates {(3,0.59)};
\addplot[blue,forget plot]coordinates {(4,0.73)};
\addplot[blue,forget plot]coordinates {(5,0.83)};
\addlegendentry{Discrete HMM}
\addlegendentry{GMM-HMM}
\end{axis}
\end{tikzpicture}
      \\
      \multicolumn{2}{c}{\footnotesize (b) Winwebsec models}
      \\ \\[-0.25ex]
      \hspace*{0.25in}{\footnotesize Test on Zbot}
      & 
      \hspace*{0.25in}{\footnotesize Test on Winwebsec}
      \\ \\[-2.5ex]
      \begin{tikzpicture}[scale=0.6, every node/.style={scale=1.0}]
\begin{axis}[
        width  = 0.6*\textwidth,
        height = 6.75cm,
        ymin=0.0,ymax=1.0,
        ytick={0.0,0.2,0.4,0.6,0.8,1.0},
        separate axis lines,
        major x tick style = transparent,
        ybar=5*\pgflinewidth,
        every axis plot/.append style={
          ybar,
          bar width=28.0pt,
          bar shift=0pt,
          fill
        },
        ylabel = {AUC},
        xtick={1,2,3,4,5},
        x tick style={draw=none},
        xticklabels={
,
$m=2$,
$m=3$,
$m=4$,
$m=5$,
},
	y tick label style={
    		/pgf/number format/.cd,
   		fixed,
   		fixed zerofill,
    		precision=2},
        x tick label style={
		inner sep=0mm},
        nodes near coords,
        every node near coord/.style={text=black},
        every node near coord/.append style={
								   /pgf/number format/.cd,
								   fixed,
								   fixed zerofill,
								   precision=2},
        enlarge x limits=0.15,
        legend cell align=left,
        legend style={
		legend pos = south west,
                	column sep=1ex
        },
]
\addplot[red]coordinates {(1,0.71)};
\addplot[blue]coordinates {(2,0.68)};
\addplot[blue,forget plot]coordinates {(3,0.72)};
\addplot[blue,forget plot]coordinates {(4,0.66)};
\addplot[blue,forget plot]coordinates {(5,0.64)};
\addlegendentry{Discrete HMM}
\addlegendentry{GMM-HMM}
\end{axis}
\end{tikzpicture}
      &
      \begin{tikzpicture}[scale=0.6, every node/.style={scale=1.0}]
\begin{axis}[
        width  = 0.6*\textwidth,
        height = 6.75cm,
        ymin=0.0,ymax=1.0,
        ytick={0.0,0.2,0.4,0.6,0.8,1.0},
        separate axis lines,
        major x tick style = transparent,
        ybar=5*\pgflinewidth,
        every axis plot/.append style={
          ybar,
          bar width=28.0pt,
          bar shift=0pt,
          fill
        },
        ylabel = {AUC},
        xtick={1,2,3,4,5},
        x tick style={draw=none},
        xticklabels={
,
$m=2$,
$m=3$,
$m=4$,
$m=5$,
},
	y tick label style={
    		/pgf/number format/.cd,
   		fixed,
   		fixed zerofill,
    		precision=2},
        x tick label style={
		inner sep=0mm},
        nodes near coords,
        every node near coord/.style={text=black},
        every node near coord/.append style={
								   /pgf/number format/.cd,
								   fixed,
								   fixed zerofill,
								   precision=2},
        enlarge x limits=0.15,
        legend cell align=left,
        legend style={
		legend pos = south west,
                	column sep=1ex
        },
]
\addplot[red]coordinates {(1,0.61)};
\addplot[blue]coordinates {(2,0.61)};
\addplot[blue,forget plot]coordinates {(3,0.65)};
\addplot[blue,forget plot]coordinates {(4,0.61)};
\addplot[blue,forget plot]coordinates {(5,0.57)};
\addlegendentry{Discrete HMM}
\addlegendentry{GMM-HMM}
\end{axis}
\end{tikzpicture}
      \\
      \multicolumn{2}{c}{\footnotesize (c) Zeroaccess models}
      \\ \\[-0.25ex]
      \end{tabular}
      \caption{Average AUC}\label{fig:avgAUV}
\end{figure*}

Figure~\ref{fig:avgAUV} gives the average AUC (over the~5 folds) 
for models trained with discrete HMMs and the GMM-HMMs with different values 
for~$m$, the number of Gaussians in the mixture. For most of the models, 
the GMM-HMM is able to obtain comparable results to the discrete HMM,
and it does slightly outperform a discrete HMM in some cases.
but the improvement is slight.

The results in Figure~\ref{fig:avgAUV} indicate that for opcodes sequences,
GMM-HMMs perform comparably to discrete HMMs. However, GMM-HMMs
are more complex and more challenging to train, and the additional 
complexity does not appear to be warranted in this case. But, this is not
surprising, as opcode sequences are inherently discrete features.
To obtain a more useful comparison, we next consider GMM-HMMs trained 
on continuous features.

\subsection{Entropy Features}\label{chap:entropy}

GMM-HMMs are designed for continuous data, as opposed to discrete features, 
such as opcodes. Thus to take full advantage of the GMM-HMM technique, 
we consider continuous entropy based features. 

We use a similar feature-extraction method as in~\cite{baysa}. Specifically, 
we consider the raw bytes of an executable file, and 
we define a window size over which we compute the entropy.
Then we slide the window by a fixed amount and repeat the entropy calculation. 
Both the window size and the slide amount are parameters that need to be tuned 
to obtain optimal performance. In general, the slide will be smaller than 
the window size to ensure no information is lost.

Entropy is computed using Shannon's well known formula~\cite{shannon} 
$$
  E = - \sum\limits_{x\in W_i}p(x)\,\log_2p(x) ,
$$ 
where~$W_i$ is the~$i^{\thth}$ window, and~$p(x)$ is the frequency of the 
occurrence of byte~$x$ within window~$W_i$. 

The entropy tends to be smoothed out with larger window sizes. We
want to select a window size sufficiently large so that we reduce
noise, but not so large as to lose useful information. 
Examples of entropy plots for different parameters
are given in Figure~\ref{fig:entropy-windowsize}. 

\begin{figure}[!htb]
     \centering
     \begin{tabular}{c}
\begin{tikzpicture}[scale=0.6]
\begin{axis}[smooth,
		   width=0.7\textwidth,
		   height=0.575\textwidth,
	 	   x tick label style={
   		 	/pgf/number format/.cd,
			/pgf/number format/1000 sep={},
   			fixed,
   			fixed zerofill,
    			precision=0
		   },
	 	   y tick label style={
    		 	/pgf/number format/.cd,
   			fixed,
   			fixed zerofill,
    			precision=0
		    },
                    xmin=-7,xmax=362,
                    ymin=0,ymax=7.75,
                    xtick={0,50,100,150,200,250,300,350},
                    ytick={0,1,2,3,4,5,6,7},
                    xlabel={Window number},
                    ylabel={Entropy}] 
\addplot[color=blue,ultra thick,
no marks] coordinates {
(1,2.232220085)
(2,1.333499723)
(3,1.934922204)
(4,4.246044196)
(5,4.900899385)
(6,4.740226408)
(7,5.932202928)
(8,6.534664402)
(9,6.729316594)
(10,6.590108011)
(11,6.463380998)
(12,6.51602601)
(13,6.618684334)
(14,6.675996618)
(15,6.738591898)
(16,6.615360144)
(17,6.531480287)
(18,6.662379419)
(19,6.635078412)
(20,6.616830536)
(21,6.574054953)
(22,6.416789405)
(23,6.464875674)
(24,6.568682859)
(25,6.624122083)
(26,6.786282933)
(27,6.758727908)
(28,6.809452592)
(29,6.69636661)
(30,6.704332595)
(31,6.673646978)
(32,6.81339375)
(33,6.812700959)
(34,6.558485963)
(35,6.630000188)
(36,6.783717423)
(37,6.748190585)
(38,6.594277563)
(39,6.77528296)
(40,6.82284449)
(41,6.67166604)
(42,6.541199018)
(43,6.421861432)
(44,6.655927716)
(45,6.713530403)
(46,6.56212083)
(47,6.655086155)
(48,6.696172365)
(49,6.670977308)
(50,6.605840288)
(51,6.647188457)
(52,6.573471433)
(53,6.56842162)
(54,6.600574312)
(55,6.598028636)
(56,6.639202738)
(57,6.562771938)
(58,6.578796068)
(59,6.6685601)
(60,6.658344873)
(61,6.661181805)
(62,6.538615769)
(63,6.532243928)
(64,6.60584061)
(65,6.582313919)
(66,6.730329791)
(67,6.699874044)
(68,6.619252556)
(69,6.697953826)
(70,6.757572157)
(71,6.823207527)
(72,6.582456558)
(73,6.47023089)
(74,6.670921417)
(75,6.730703591)
(76,6.558503262)
(77,6.497456157)
(78,6.667283909)
(79,6.634179174)
(80,6.694216002)
(81,6.633678348)
(82,6.661141457)
(83,6.59232266)
(84,6.659419698)
(85,6.778940157)
(86,6.748871615)
(87,6.495381057)
(88,6.670011216)
(89,6.706580498)
(90,6.646867584)
(91,6.571852714)
(92,6.481505359)
(93,6.632401049)
(94,6.693528666)
(95,6.657656028)
(96,6.682853456)
(97,6.693091142)
(98,6.615930534)
(99,6.730667171)
(100,6.665549225)
(101,6.635263009)
(102,6.849167199)
(103,6.830920648)
(104,6.77580906)
(105,6.706989456)
(106,6.591800613)
(107,6.573706454)
(108,6.565960596)
(109,6.652510368)
(110,6.805912135)
(111,6.764405615)
(112,6.630060445)
(113,6.537753099)
(114,6.74339755)
(115,6.73731102)
(116,6.566989074)
(117,6.42113804)
(118,6.413961488)
(119,6.453767207)
(120,6.361293829)
(121,6.604009432)
(122,6.595659148)
(123,6.539943932)
(124,6.59336955)
(125,6.469858079)
(126,6.421738966)
(127,6.508825045)
(128,6.497685489)
(129,6.617957162)
(130,6.755292133)
(131,6.715760275)
(132,6.355424115)
(133,6.455935749)
(134,6.701969305)
(135,6.769759745)
(136,6.799489779)
(137,6.657421909)
(138,6.594464532)
(139,6.665736228)
(140,6.597019141)
(141,6.593359799)
(142,6.566357755)
(143,6.657264558)
(144,6.785262725)
(145,6.656117322)
(146,6.628229356)
(147,6.59233021)
(148,6.522016097)
(149,6.732393785)
(150,6.738491753)
(151,6.780756559)
(152,6.649090485)
(153,6.477348659)
(154,6.514671765)
(155,6.548674207)
(156,6.683554621)
(157,6.80570303)
(158,6.744878956)
(159,6.662858361)
(160,6.756109087)
(161,6.913657539)
(162,6.768588727)
(163,6.656431408)
(164,6.696647862)
(165,6.525272682)
(166,6.670236733)
(167,6.620965183)
(168,6.487554339)
(169,6.617674779)
(170,6.661780773)
(171,6.629645955)
(172,6.615103945)
(173,6.658533538)
(174,6.56908022)
(175,6.517143037)
(176,6.563251901)
(177,6.499731508)
(178,6.701705709)
(179,6.811799966)
(180,6.675172664)
(181,6.469128083)
(182,6.567929714)
(183,6.637234579)
(184,6.758839835)
(185,6.749294985)
(186,6.71103468)
(187,6.716846311)
(188,6.661227221)
(189,6.711672614)
(190,6.637125304)
(191,6.607302889)
(192,6.626013537)
(193,6.628523122)
(194,6.585914781)
(195,6.626154019)
(196,6.514164986)
(197,6.371202307)
(198,6.410494374)
(199,6.225066112)
(200,6.091722222)
(201,6.184901913)
(202,6.471083895)
(203,6.401472242)
(204,6.330074096)
(205,6.418608286)
(206,6.335216041)
(207,6.366110892)
(208,6.48433384)
(209,6.459891142)
(210,6.664285422)
(211,6.744231479)
(212,6.597867653)
(213,6.736589231)
(214,6.804034678)
(215,6.796726678)
(216,6.68690492)
(217,6.554297269)
(218,6.524221343)
(219,6.433603413)
(220,6.584762336)
(221,6.625843094)
(222,6.619410078)
(223,6.690172278)
(224,6.880751789)
(225,7.017357159)
(226,6.778625047)
(227,6.640893015)
(228,6.695792984)
(229,6.643408837)
(230,6.128379946)
(231,5.350262469)
(232,4.378419726)
(233,5.391624029)
(234,6.654399151)
(235,6.626704298)
(236,6.603120621)
(237,6.376989697)
(238,6.569820593)
(239,6.460658925)
(240,5.961754799)
(241,6.294960618)
(242,6.501908596)
(243,5.928383321)
(244,6.013153337)
(245,6.169897857)
(246,6.398749251)
(247,6.433744646)
(248,6.394031427)
(249,6.457398554)
(250,6.219821664)
(251,6.317518355)
(252,6.736714689)
(253,6.452902929)
(254,6.38594652)
(255,6.634514436)
(256,6.545893911)
(257,6.484246907)
(258,6.300293289)
(259,6.610112881)
(260,6.734274692)
(261,6.42870168)
(262,6.201770253)
(263,5.602680956)
(264,4.380578558)
(265,4.759600831)
(266,5.526460857)
(267,5.280179081)
(268,5.030619757)
(269,4.956495563)
(270,5.208547775)
(271,5.113128542)
(272,5.147713356)
(273,5.315334052)
(274,5.337723003)
(275,5.560013496)
(276,5.478311243)
(277,5.735959231)
(278,5.759784569)
(279,6.271062014)
(280,6.883737151)
(281,6.912320028)
(282,6.713103255)
(283,6.356784575)
(284,5.444922833)
(285,5.749916126)
(286,6.231251528)
(287,6.198982968)
(288,6.523113768)
(289,6.569679012)
(290,6.525243144)
(291,6.396536754)
(292,6.220455956)
(293,6.334635863)
(294,5.652265663)
(295,3.425346867)
(296,4.144958444)
(297,4.12337786)
(298,3.738670027)
(299,4.453253261)
(300,4.718043828)
(301,5.600528412)
(302,5.952954397)
(303,6.133778769)
(304,6.145750875)
(305,5.533416613)
(306,4.046240513)
(307,3.963659727)
(308,3.19277672)
(309,1.247251539)
(310,2.576708023)
(311,4.685719255)
(312,4.609517438)
(313,4.634545798)
(314,4.522500372)
(315,4.514495502)
(316,4.604925464)
(317,4.824338181)
(318,5.77250266)
(319,6.221440285)
(320,4.817805732)
(321,1.752659624)
(322,0.859778666)
(323,0.75427541)
(324,0.272970858)
(325,0.272970858)
(326,0.272970858)
(327,0.272970858)
(328,0.272970858)
(329,0.272970858)
(330,1.514936467)
(331,2.417513302)
(332,2.881275759)
(333,3.660377271)
(334,4.43036207)
(335,4.342184418)
(336,3.422109142)
(337,2.293757309)
(338,2.732114876)
(339,3.28091609)
(340,3.70625308)
(341,2.967418466)
(342,1.569006943)
(343,2.893122253)
(344,3.222085551)
(345,2.843141653)
(346,4.388408381)
(347,5.152390756)
(348,5.337515096)
(349,6.677871984)
(350,6.974171704)
(351,3.064050687)
(352,3.044608085)
(353,6.697979289)
(354,7.541051495)
(355,6.62830148)
};
\end{axis}
\end{tikzpicture} \\
         (a) Window size = 512 \\
         \\
     	 \input figures/entropy_128.tex \\
         (b) Window size = 128
         \\ \\[-2ex]
         \end{tabular}
        \caption{Entropy plots}\label{fig:entropy-windowsize}
\end{figure}

Based on the results in~\cite{baysa}, we use half of the window size as the slide amount. 
To select the best values for the parameters, we conduct experiments 
with the window and slide combinations listed in Table~\ref{table:window_size}.
Also, as part of the parameter tuning process, we selected~$\epsilon$ 
in~\eref{eqa:continous-probability} to 
be~$0.000001$ for both Zbot and Zeroaccess,
while we find~$0.1$ is optimal for Winwebsec.

\begin{table}
\centering
\caption{Window size and slide amount}\label{table:window_size}
\begin{tabular}{ c|ccc } 
 \midrule\midrule
 \textbf{Window size} & 512 & 256 & 128 \\ \midrule
 \textbf{Slide} & 256 & 128 & 64\\ 
 \midrule\midrule
\end{tabular}
\end{table}

For models trained on Zbot, the results of our experiments with the
different window and slide size pairings 
in Table~\ref{table:window_size}
are given in Figure~\ref{fig:roc-windowsize}.
The corresponding bar graphs for models
trained on Winwebsec and Zeroaccess are given
in Figures~\ref{fig:roc-windowsize-winwebsec} 
and~\ref{fig:roc-windowsize-zeroaccess}, respectively,
which can be found in the Appendix.
Note that we have experimented with the number
of Gaussians in our mixture ranging from~$m=2$ to~$m=5$.
We see that a window size of size~512 performs the 
worst, while window sizes of size~256 and~128
give improved results, with size~128 being slightly better
than~256. The optimal number of Gaussians
depends on the families we are classifying.
The results of analogous experiments training models on 
Winwebsec and Zeroaccess are given in Figures~\ref{fig:roc-windowsize-winwebsec}
and~\ref{fig:roc-windowsize-zeroaccess}, respectively.

\begin{figure}[!htb]
      \centering
      \begin{tabular}{cc}
      {\scriptsize\hspace*{0.25in} Test on Winwebsec}
      & 
      {\scriptsize\hspace*{0.25in} Test on Zeroaccess} \\[-0.25ex]
\begin{tikzpicture}[scale=0.6, every node/.style={scale=1.0}]
    \begin{axis}[
        width  = 0.37*\textwidth,
        height = 5.25cm,
        ymin=0,ymax=1.14,
        ytick={0.0,0.2,0.4,0.6,0.8,1.0},
        major x tick style = transparent,
        ybar=5*\pgflinewidth,
        bar width=22pt,
        xlabel = {Number of mixture components},
        ylabel = {AUC},
        xlabel shift = 6pt,
        symbolic x coords={
$m=2$,
$m=3$,
$m=4$,
$m=5$
	},
        xtick={
$m=2$,
$m=3$,
$m=4$,
$m=5$
	},
	y tick label style={
		font=\small,
    		/pgf/number format/.cd,
   		fixed,
   		fixed zerofill,
		1000 sep={},
    		precision=2},
        x tick label style={
		font=\footnotesize,
		inner sep=0mm
		},
        nodes near coords,
        every node near coord/.append style={
								   font=\footnotesize,
								   /pgf/number format/.cd,
								   	fixed zerofill,
									1000 sep={},
									precision=2
								   },
        enlarge x limits=0.18,
    ]
\addplot[fill=blue,opacity=1.00] 
coordinates {
($m=2$, 0.92)
($m=3$, 0.91)
($m=4$, 0.92)
($m=5$, 0.91)
};
\end{axis}
\end{tikzpicture}
      &
\begin{tikzpicture}[scale=0.6, every node/.style={scale=1.0}]
    \begin{axis}[
        width  = 0.37*\textwidth,
        height = 5.25cm,
        ymin=0,ymax=1.14,
        ytick={0.0,0.2,0.4,0.6,0.8,1.0},
        major x tick style = transparent,
        ybar=5*\pgflinewidth,
        bar width=22pt,
        xlabel = {Number of mixture components},
        ylabel = {AUC},
        xlabel shift = 6pt,
        symbolic x coords={
$m=2$,
$m=3$,
$m=4$,
$m=5$
	},
        xtick={
$m=2$,
$m=3$,
$m=4$,
$m=5$
	},
	y tick label style={
		font=\small,
    		/pgf/number format/.cd,
   		fixed,
   		fixed zerofill,
		1000 sep={},
    		precision=2},
        x tick label style={
		font=\footnotesize,
		inner sep=0mm
		},
        nodes near coords,
        every node near coord/.append style={
								   font=\footnotesize,
								   /pgf/number format/.cd,
								   	fixed zerofill,
									1000 sep={},
									precision=2
								   },
        enlarge x limits=0.18,
    ]
\addplot[fill=blue,opacity=1.00] 
coordinates {
($m=2$, 0.86)
($m=3$, 0.87)
($m=4$, 0.87)
($m=5$, 0.88)
};
\end{axis}
\end{tikzpicture} \\
      \multicolumn{2}{c}{(a) $\mbox{Window size} = 512$} \\ \\[-2ex]
      {\scriptsize\hspace*{0.25in} Test on Winwebsec}
      & 
      {\scriptsize\hspace*{0.25in} Test on Zeroaccess} \\[-0.25ex]
\begin{tikzpicture}[scale=0.6, every node/.style={scale=1.0}]
    \begin{axis}[
        width  = 0.37*\textwidth,
        height = 5.25cm,
        ymin=0,ymax=1.14,
        ytick={0.0,0.2,0.4,0.6,0.8,1.0},
        major x tick style = transparent,
        ybar=5*\pgflinewidth,
        bar width=22pt,
        xlabel = {Number of mixture components},
        ylabel = {AUC},
        xlabel shift = 6pt,
        symbolic x coords={
$m=2$,
$m=3$,
$m=4$,
$m=5$
	},
        xtick={
$m=2$,
$m=3$,
$m=4$,
$m=5$
	},
	y tick label style={
		font=\small,
    		/pgf/number format/.cd,
   		fixed,
   		fixed zerofill,
		1000 sep={},
    		precision=2},
        x tick label style={
		font=\footnotesize,
		inner sep=0mm
		},
        nodes near coords,
        every node near coord/.append style={
								   font=\footnotesize,
								   /pgf/number format/.cd,
								   	fixed zerofill,
									1000 sep={},
									precision=2
								   },
        enlarge x limits=0.18,
    ]
\addplot[fill=blue,opacity=1.00] 
coordinates {
($m=2$, 0.846)
($m=3$, 0.936)
($m=4$, 0.968)
($m=5$, 0.856)
};
\end{axis}
\end{tikzpicture}
      &
\begin{tikzpicture}[scale=0.6, every node/.style={scale=1.0}]
    \begin{axis}[
        width  = 0.37*\textwidth,
        height = 5.25cm,
        ymin=0,ymax=1.14,
        ytick={0.0,0.2,0.4,0.6,0.8,1.0},
        major x tick style = transparent,
        ybar=5*\pgflinewidth,
        bar width=22pt,
        xlabel = {Number of mixture components},
        ylabel = {AUC},
        xlabel shift = 6pt,
        symbolic x coords={
$m=2$,
$m=3$,
$m=4$,
$m=5$
	},
        xtick={
$m=2$,
$m=3$,
$m=4$,
$m=5$
	},
	y tick label style={
		font=\small,
    		/pgf/number format/.cd,
   		fixed,
   		fixed zerofill,
		1000 sep={},
    		precision=2},
        x tick label style={
		font=\footnotesize,
		inner sep=0mm
		},
        nodes near coords,
        every node near coord/.append style={
								   font=\footnotesize,
								   /pgf/number format/.cd,
								   	fixed zerofill,
									1000 sep={},
									precision=2
								   },
        enlarge x limits=0.18,
    ]
\addplot[fill=blue,opacity=1.00] 
coordinates {
($m=2$, 0.938)
($m=3$, 0.924)
($m=4$, 0.938)
($m=5$, 0.918)
};
\end{axis}
\end{tikzpicture} \\
      \multicolumn{2}{c}{(b) $\mbox{Window size} = 256$} \\ \\[-2ex]
      {\scriptsize\hspace*{0.25in} Test on Winwebsec}
      & 
      {\scriptsize\hspace*{0.25in} Test on Zeroaccess} \\[-0.25ex]
\begin{tikzpicture}[scale=0.6, every node/.style={scale=1.0}]
    \begin{axis}[
        width  = 0.37*\textwidth,
        height = 5.25cm,
        ymin=0,ymax=1.14,
        ytick={0.0,0.2,0.4,0.6,0.8,1.0},
        major x tick style = transparent,
        ybar=5*\pgflinewidth,
        bar width=22pt,
        xlabel = {Number of mixture components},
        ylabel = {AUC},
        xlabel shift = 6pt,
        symbolic x coords={
$m=2$,
$m=3$,
$m=4$,
$m=5$
	},
        xtick={
$m=2$,
$m=3$,
$m=4$,
$m=5$
	},
	y tick label style={
		font=\small,
    		/pgf/number format/.cd,
   		fixed,
   		fixed zerofill,
		1000 sep={},
    		precision=2},
        x tick label style={
		font=\footnotesize,
		inner sep=0mm
		},
        nodes near coords,
        every node near coord/.append style={
								   font=\footnotesize,
								   /pgf/number format/.cd,
								   	fixed zerofill,
									1000 sep={},
									precision=2
								   },
        enlarge x limits=0.18,
    ]
\addplot[fill=blue,opacity=1.00] 
coordinates {
($m=2$, 0.892)
($m=3$, 0.966)
($m=4$, 0.976)
($m=5$, 0.954)
};
\end{axis}
\end{tikzpicture}
      &
\begin{tikzpicture}[scale=0.6, every node/.style={scale=1.0}]
    \begin{axis}[
        width  = 0.37*\textwidth,
        height = 5.25cm,
        ymin=0,ymax=1.14,
        ytick={0.0,0.2,0.4,0.6,0.8,1.0},
        major x tick style = transparent,
        ybar=5*\pgflinewidth,
        bar width=22pt,
        xlabel = {Number of mixture components},
        ylabel = {AUC},
        xlabel shift = 6pt,
        symbolic x coords={
$m=2$,
$m=3$,
$m=4$,
$m=5$
	},
        xtick={
$m=2$,
$m=3$,
$m=4$,
$m=5$
	},
	y tick label style={
		font=\small,
    		/pgf/number format/.cd,
   		fixed,
   		fixed zerofill,
		1000 sep={},
    		precision=2},
        x tick label style={
		font=\footnotesize,
		inner sep=0mm
		},
        nodes near coords,
        every node near coord/.append style={
								   font=\footnotesize,
								   /pgf/number format/.cd,
								   	fixed zerofill,
									1000 sep={},
									precision=2
								   },
        enlarge x limits=0.18,
    ]
\addplot[fill=blue,opacity=1.00] 
coordinates {
($m=2$, 0.936)
($m=3$, 0.918)
($m=4$, 0.928)
($m=5$, 0.922)
};
\end{axis}
\end{tikzpicture} \\
      \multicolumn{2}{c}{(c) $\mbox{Window size} = 128$} \\ \\[-2ex]
      \end{tabular}
   \caption{Entropy vs window size for Zbot models}\label{fig:roc-windowsize}
\end{figure}


In Table~\ref{tab:best_compare}, we provide a direct comparison of discrete HMMs
trained on opcodes to GMM-HMM trained on opcodes and
the best GMM-HMM models trained on entropy sequences.
In every case, the entropy-trained GMM-HMM  
outperforms the corresponding opcode based models.
It is also worth noting that computing an entropy sequence is more efficient
than extracting mnemonic opcodes. While it is costlier to train a GMM-HMM,
the scoring cost is similar to a discrete HMM. Since training is one-time work,
efficiency considerations also favor entropy-based GMM-HMMs.


\begin{table}[!htb]
\centering
\caption{Comparison of discrete HMM and GMM-HMM}\label{tab:best_compare}
\resizebox{0.475\textwidth}{!}{
\begin{tabular}{cc|cccc} 
 \midrule\midrule
 \multirow{2}{*}{\textbf{Train}} &	\multirow{2}{*}{\textbf{Test}} & \textbf{Opcode} & \textbf{Opcode} & \textbf{Entropy} \\
 	& & \textbf{HMM} & \textbf{GMM-HMM} & \textbf{GMM-HMM} \\
 \midrule
 Zbot & Zeroaccess & 0.87 & 0.90 & \textbf{0.94} \\
 Zbot & Winwebsec & 0.69 & 0.69 & \textbf{0.98} \\ 
 Zeroaccess & Zbot & 0.71 & 0.72 & \textbf{0.77} \\ 
 Zeroaccess & Winwebsec & 0.61 & 0.65 & \textbf{0.99} \\ 
 Winwebsec & Zbot & 0.72 & 0.66 & \textbf{1.00} \\ 
 Winwebsec & Zeroaccess & 0.83 & 0.83 & \textbf{1.00} \\ 
 \midrule\midrule
\end{tabular}
}
\end{table}

From Table~\ref{tab:best_compare} we see that 
GMM-HMMs trained on entropy perform dramatically better than
discrete HMMs, except in the two cases where
models where Zbot and Zeroaccess are involved.
To gain further insight into these anomalous case, we use the
Kullback–Leibler (KL) divergence~\cite{kldivergence} to 
compare the probability distributions defined by of our trained GMM-HMM
models. The KL divergence between two probability distributions is given by
\begin{equation}\label{eqa:kl-divergence}
    \mbox{KL}(p\,\|\,q) = \int_{-\infty}^{\infty} \!\! p(x)\,\log\,\frac{p(x)}{q(x)} ,
\end{equation}
where~$p$ and~$q$ are probability density functions. 
Note that the KL divergence in~\eref{eqa:kl-divergence} is not symmetric, 
and hence not a true distance measure.
We compute a symmetric version of the divergence 
for models~${\cal M}_1$ and~${\cal M}_2$ as
\begin{equation}\label{eqa:kl-divergence-sym}
  \mbox{KL}({\cal M}_1, {\cal M}_2) 
   	= \frac{\mbox{KL}({\cal M}_1\,\|\,{\cal M}_2) 
  	+ \mbox{KL}({\cal M}_2\,\|\,{\cal M}_1)}{2} .
\end{equation}

Using equation~\eref{eqa:kl-divergence-sym} we obtain the 
(symmetric) divergence results in Table~\ref{tab:modeldivergence}.
We see that the Zbot and Zeroaccess models are
much closer in terms of KL divergence, as compared to the 
other two pairs. Thus we would expect
GMM-HMM models to have more difficulty 
distinguishing these two families from each other,
as compared to the models generated for the other pairs of families.

\begin{table}[!htb]
\centering
\caption{The KL divergence of different models}\label{tab:modeldivergence}
\begin{tabular}{ c|c} 
 \midrule\midrule
 \textbf{Models} & \textbf{KL divergence}\\ 
 \midrule
 Zbot, Zeroaccess & \z611.58 \\
 Zbot, Winwebsec & 1594.05\\ 
 Zeroaccess, Winwebsec & 1524.39\\ 
 \midrule\midrule
\end{tabular}
\end{table}

Curiously, the models trained on Zbot and tested on Zeroaccess 
perform well.\footnote{It is worth noting that the opcode based models also performed well in this case.}
Hence, a relatively small KL divergence does
not rule out the possibility that models can be useful, but 
intuitively, a large divergence would seem to be an 
indicator of potentially challenging cases. This issue requires further study.

\section{Conclusion and Future Work\label{chap:result}}

In this paper, we have explored the usage of GMM-HMMs for 
malware classification, We compared GMM-HMMs to discrete HMMs 
using opcode sequences, and we further experimented with entropy sequences
as features for GMM-HMMs. With the opcode sequence features, we
were able to obtained results with GMM-HMMs that are comparable to
those obtained using discrete HMMs, However, we expect GMM-HMMs 
to perform best on features that are naturally continuous,
so we also experimented with byte-based entropy sequences.
In this latter set of experiments, the GMM-HMM technique
yielded stronger results than the discrete HMM in all cases---and in four
of the six cases, the improvement was large. We also directly compared
the GMMs of our trained models using KL divergence,
which seems to provide insight into the most challenging cases.

For future work, more extensive experiments over larger numbers 
of families with larger numbers of samples per family would be valuable. 
True multiclass experiments based on GMM-HMM scores would also
be of interest. Further analysis of the KL divergence of GMM-HMMs
might provide useful insights into these models.

\bibliographystyle{apalike}

{\small
\bibliography{references.bib}

\begin{thebibliography}{}

\bibitem[Alfakih et~al., 2020]{ALFAKIH2020101218}
Alfakih, M., Keche, M., Benoudnine, H., and Meche, A. (2020).
\newblock Improved {G}aussian mixture modeling for accurate {Wi-Fi} based
  indoor localization systems.
\newblock {\em Physical Communication}, 43.

\bibitem[Bansal et~al., 2008]{bansal2008improved}
Bansal, P., Kant, A., Kumar, S., Sharda, A., and Gupta, S. (2008).
\newblock Improved hybrid model of hmm/gmm for speech recognition.
\newblock In {\em International Conference on Intelligent Information and
  Engineering Systems}, INFOS 2008.

\bibitem[Baysa et~al., 2013]{baysa}
Baysa, D., Low, R., and Stamp, M. (2013).
\newblock Structural entropy and metamorphic malware.
\newblock {\em Journal of Computer Virology and Hacking Techniques},
  9(4):179--192.

\bibitem[Ben Abdel~Ouahab et~al., 2020]{knn}
Ben Abdel~Ouahab, I., Bouhorma, M., Boudhir, A.~A., and El~Aachak, L. (2020).
\newblock Classification of grayscale malware images using the $k$-nearest
  neighbor algorithm.
\newblock In Ben~Ahmed, M., Boudhir, A.~A., Santos, D., El~Aroussi, M., and
  Karas, {\.{I}}.~R., editors, {\em Innovations in Smart Cities Applications},
  pages 1038--1050. Springer, 3 edition.

\bibitem[Bradley, 1997]{rocBrad}
Bradley, A.~P. (1997).
\newblock The use of the area under the {ROC} curve in the evaluation of
  machine learning algorithms.
\newblock {\em Pattern Recognition}, 30(7):1145--1159.

\bibitem[Brown Corpus of standard American English, 1961]{english-text}
Brown Corpus of standard American English (1961).
\newblock The {B}rown corpus of standard {A}merican {E}nglish.
\newblock \url{http://www.cs.toronto.edu/ ~gpenn/csc401/a1res.html}.

\bibitem[Cave and Neuwirth, 1980]{cave}
Cave, R.~L. and Neuwirth, L.~P. (1980).
\newblock Hidden {M}arkov models for {E}nglish.
\newblock In Ferguson, J.~D., editor, {\em Hidden {M}arkov Models for Speech}.
  IDA-CCR.

\bibitem[Chen and Wu, 2019]{CHEN20199}
Chen, Y. and Wu, W. (2019).
\newblock Separation of geochemical anomalies from the sample data of unknown
  distribution population using gaussian mixture model.
\newblock {\em Computers \&\ Geosciences}, 125:9--18.

\bibitem[Dang et~al., 2017]{DANG201787}
Dang, S., Chaudhury, S., Lall, B., and Roy, P.~K. (2017).
\newblock Learning effective connectivity from {fMRI} using autoregressive
  hidden {M}arkov model with missing data.
\newblock {\em Journal of Neuroscience Methods}, 278:87--100.

\bibitem[Fraley and Raftery, 2002]{gmm}
Fraley, C. and Raftery, A.~E. (2002).
\newblock Model-based clustering, discriminant analysis, and density
  estimation.
\newblock {\em Journal of the American Statistical Association},
  97(458):611--631.

\bibitem[Gallop, 2006]{gallopj}
Gallop, J. (2006).
\newblock Facies probability from mixture distributions with non‐stationary
  impedance errors.
\newblock In {\em {SEG} Technical Program Expanded Abstracts 2006}, pages
  1801--1805. Society of Exploration Geophysicists.

\bibitem[Gao et~al., 2020]{GAO2020107815}
Gao, Z., Sun, Z., and Liang, S. (2020).
\newblock Probability density function for wave elevation based on {G}aussian
  mixture models.
\newblock {\em Ocean Engineering}, 213.

\bibitem[{Guoning Hu} and {DeLiang Wang}, 2004]{1333078}
{Guoning Hu} and {DeLiang Wang} (2004).
\newblock Monaural speech segregation based on pitch tracking and amplitude
  modulation.
\newblock {\em IEEE Transactions on Neural Networks}, 15(5):1135--1150.

\bibitem[Interrante-Grant and Kaeli, 2018]{interrantegaussian}
Interrante-Grant, A.~M. and Kaeli, D. (2018).
\newblock Gaussian mixture models for dynamic malware clustering.
\newblock
  \url{https://coe.northeastern.edu/wp-content/uploads/pdfs/coe/research/embark/4-interrante-grant.alex_final.pdf}.

\bibitem[Joyce, 2011]{kldivergence}
Joyce, J.~M. (2011).
\newblock {K}ullback-{L}eibler divergence.
\newblock In Lovric, M., editor, {\em International Encyclopedia of Statistical
  Science}, pages 720--722. Springer.

\bibitem[Juang, 1985]{6769559}
Juang, B. (1985).
\newblock Maximum-likelihood estimation for mixture multivariate stochastic
  observations of {M}arkov chains.
\newblock {\em AT\&T Technical Journal}, 64(6):1235--1249.

\bibitem[{Kalash} et~al., 2018]{nnmalware}
{Kalash}, M., {Rochan}, M., {Mohammed}, N., {Bruce}, N. D.~B., {Wang}, Y., and
  {Iqbal}, F. (2018).
\newblock Malware classification with deep convolutional neural networks.
\newblock In {\em 2018 9th IFIP International Conference on New Technologies,
  Mobility and Security}, NTMS, pages 1--5.

\bibitem[Kruczkowski and Szynkiewicz, 2014]{svm}
Kruczkowski, M. and Szynkiewicz, E.~N. (2014).
\newblock Support vector machine for malware analysis and classification.
\newblock In {\em 2014 IEEE/WIC/ACM International Joint Conferences on Web
  Intelligence and Intelligent Agent Technologies}, WI-IAT '14, pages 415--420.

\bibitem[Laraba and Tilmanne, 2016]{hmmdancing}
Laraba, S. and Tilmanne, J. (2016).
\newblock Dance performance evaluation using hidden {M}arkov models.
\newblock {\em Computer Animation and Virtual Worlds}, 27(3-4):321--329.

\bibitem[McLachlan and Peel, 2004]{mclachlan2004finite}
McLachlan, G. and Peel, D. (2004).
\newblock {\em Finite Mixture Models}.
\newblock Wiley.

\bibitem[Milosevic, 2013]{malwarehistory}
Milosevic, N. (2013).
\newblock History of malware.
\newblock \url{https://arxiv.org/abs/1302.5392}.

\bibitem[Nappa et~al., 2015]{malicia-dataset}
Nappa, A., Rafique, M.~Z., and Caballero, J. (2015).
\newblock The {MALICIA} dataset: Identification and analysis of drive-by
  download operations.
\newblock {\em International Journal of Information Security}, 14(1):15--33.

\bibitem[Neville and Gibb, 2013]{zeroacce66:online}
Neville, A. and Gibb, R. (2013).
\newblock {ZeroAccess} {Indepth}.
\newblock \url{https://docs.broadcom.com/doc/zeroaccess-indepth-13-en}.

\bibitem[Nguyen, 2016]{hmmcon}
Nguyen, L. (2016).
\newblock Continuous observation hidden {M}arkov model.
\newblock {\em Revista Kasmera}, 44(6):65--149.

\bibitem[Qiao et~al., 2019]{QIAO2019104628}
Qiao, J., Cai, X., Xiao, Q., Chen, Z., Kulkarni, P., Ferris, C., Kamarthi, S.,
  and Sridhar, S. (2019).
\newblock Data on {MRI} brain lesion segmentation using $k$-means and
  {G}aussian mixture model-expectation maximization.
\newblock {\em Data in Brief}, 27.

\bibitem[Rabiner, 1989]{rabiner}
Rabiner, L.~R. (1989).
\newblock A tutorial on hidden {M}arkov models and selected applications in
  speech recognition.
\newblock {\em Proceedings of the IEEE}, 77(2):257--286.

\bibitem[Raitoharju et~al., 2020]{RAITOHARJU2020107330}
Raitoharju, M., García-Fern\'{a}ndez, A., Hostettler, R., Pich\'{e}, R., and
  S{\"{a}}rkk{\"{a}}, S. (2020).
\newblock Gaussian mixture models for signal mapping and positioning.
\newblock {\em Signal Processing}, 168:107330.

\bibitem[Reynolds, 2015]{reynolds2009gaussian}
Reynolds, D. (2015).
\newblock Gaussian mixture models.
\newblock In Li, S.~Z. and Jain, A.~K., editors, {\em Encyclopedia of
  Biometrics}, pages 827--832. Springer.

\bibitem[Stamp, 2018]{stamp}
Stamp, M. (2018).
\newblock A revealing introduction to hidden {M}arkov models.
\newblock \url{https://www.cs.sjsu.edu/~stamp/RUA/HMM.pdf}.

\bibitem[{Stanculescu} et~al., 2014]{6680664}
{Stanculescu}, I., {Williams}, C. K.~I., and {Freer}, Y. (2014).
\newblock Autoregressive hidden {M}arkov models for the early detection of
  neonatal sepsis.
\newblock {\em IEEE Journal of Biomedical and Health Informatics},
  18(5):1560--1570.

\bibitem[Togneri and DeSilva, 2003]{shannon}
Togneri, R. and DeSilva, C. J.~S. (2003).
\newblock {\em Fundamentals of Information Theory and Coding Design}.
\newblock CRC Press.

\bibitem[Truong and Zaharia, 2017]{hmm3dgestures}
Truong, A. and Zaharia, T. (2017).
\newblock Laban movement analysis and hidden {M}arkov models for dynamic {3D}
  gesture recognition.
\newblock {\em EURASIP Journal on Image and Video Processing}, 2017.

\bibitem[Winwebsec, 2017]{Win32Win35:online}
Winwebsec (2017).
\newblock Win32/winwebsec threat description - {M}icrosoft security
  intelligence.
\newblock
  \url{https://www.microsoft.com/en-us/wdsi/threats/malware-encyclopedia-description?Name=Win32/Winwebsec}.

\bibitem[Yao et~al., 2020]{YAO2020102711}
Yao, Z., Ge, J., Wu, Y., Lin, X., He, R., and Ma, Y. (2020).
\newblock Encrypted traffic classification based on {G}aussian mixture models
  and hidden {M}arkov models.
\newblock {\em Journal of Network and Computer Applications}, 166.

\bibitem[Zbot, 2017]{PWSWin3261:online}
Zbot (2017).
\newblock Pws:win32/zbot threat description - {M}icrosoft security
  intelligence.
\newblock
  \url{https://www.microsoft.com/en-us/wdsi/threats/malware-encyclopedia-description?Name=PWS%3AWin32%2FZbot}.

\bibitem[Zhang et~al., 2020]{ZHANG2020106603}
Zhang, F., Han, S., Gao, H., and Wang, T. (2020).
\newblock A {G}aussian mixture based hidden {M}arkov model for motion
  recognition with {3D} vision device.
\newblock {\em Computers \&\ Electrical Engineering}, 83.

\end{thebibliography}
}

\section*{\uppercase{Appendix}}

\noindent Here, we bar graphs analogous to those in 
Figure~\ref{fig:roc-windowsize} for models trained on 
Winwebsec and Zeroaccess.

\begin{figure}[!htb]
      \centering
      \begin{tabular}{cc}
      {\scriptsize\hspace*{0.25in} Test on Zbot}
      & 
      {\scriptsize\hspace*{0.25in} Test on Zeroaccess} \\[-0.25ex]
\begin{tikzpicture}[scale=0.6, every node/.style={scale=1.0}]
    \begin{axis}[
        width  = 0.37*\textwidth,
        height = 5.25cm,
        ymin=0,ymax=1.14,
        ytick={0.0,0.2,0.4,0.6,0.8,1.0},
        major x tick style = transparent,
        ybar=5*\pgflinewidth,
        bar width=22pt,
        xlabel = {Number of mixture components},
        ylabel = {AUC},
        xlabel shift = 6pt,
        symbolic x coords={
$m=2$,
$m=3$,
$m=4$,
$m=5$
	},
        xtick={
$m=2$,
$m=3$,
$m=4$,
$m=5$
	},
	y tick label style={
		font=\small,
    		/pgf/number format/.cd,
   		fixed,
   		fixed zerofill,
		1000 sep={},
    		precision=2},
        x tick label style={
		font=\footnotesize,
		inner sep=0mm
		},
        nodes near coords,
        every node near coord/.append style={
								   font=\footnotesize,
								   /pgf/number format/.cd,
								   	fixed zerofill,
									1000 sep={},
									precision=2
								   },
        enlarge x limits=0.18,
    ]
\addplot[fill=blue,opacity=1.00] 
coordinates {
($m=2$, 0.99)
($m=3$, 0.99)
($m=4$, 1.00)
($m=5$, 1.00)
};
\end{axis}
\end{tikzpicture}
      &
\begin{tikzpicture}[scale=0.6, every node/.style={scale=1.0}]
    \begin{axis}[
        width  = 0.37*\textwidth,
        height = 5.25cm,
        ymin=0,ymax=1.14,
        ytick={0.0,0.2,0.4,0.6,0.8,1.0},
        major x tick style = transparent,
        ybar=5*\pgflinewidth,
        bar width=22pt,
        xlabel = {Number of mixture components},
        ylabel = {AUC},
        xlabel shift = 6pt,
        symbolic x coords={
$m=2$,
$m=3$,
$m=4$,
$m=5$
	},
        xtick={
$m=2$,
$m=3$,
$m=4$,
$m=5$
	},
	y tick label style={
		font=\small,
    		/pgf/number format/.cd,
   		fixed,
   		fixed zerofill,
		1000 sep={},
    		precision=2},
        x tick label style={
		font=\footnotesize,
		inner sep=0mm
		},
        nodes near coords,
        every node near coord/.append style={
								   font=\footnotesize,
								   /pgf/number format/.cd,
								   	fixed zerofill,
									1000 sep={},
									precision=2
								   },
        enlarge x limits=0.18,
    ]
\addplot[fill=blue,opacity=1.00] 
coordinates {
($m=2$, 0.99)
($m=3$, 0.99)
($m=4$, 0.99)
($m=5$, 0.99)
};
\end{axis}
\end{tikzpicture} \\
      \multicolumn{2}{c}{(a) $\mbox{Window size} = 512$} \\ \\[-2ex]
      {\scriptsize\hspace*{0.25in} Test on Zbot}
      & 
      {\scriptsize\hspace*{0.25in} Test on Zeroaccess} \\[-0.25ex]
\begin{tikzpicture}[scale=0.6, every node/.style={scale=1.0}]
    \begin{axis}[
        width  = 0.37*\textwidth,
        height = 5.25cm,
        ymin=0,ymax=1.14,
        ytick={0.0,0.2,0.4,0.6,0.8,1.0},
        major x tick style = transparent,
        ybar=5*\pgflinewidth,
        bar width=22pt,
        xlabel = {Number of mixture components},
        ylabel = {AUC},
        xlabel shift = 6pt,
        symbolic x coords={
$m=2$,
$m=3$,
$m=4$,
$m=5$
	},
        xtick={
$m=2$,
$m=3$,
$m=4$,
$m=5$
	},
	y tick label style={
		font=\small,
    		/pgf/number format/.cd,
   		fixed,
   		fixed zerofill,
		1000 sep={},
    		precision=2},
        x tick label style={
		font=\footnotesize,
		inner sep=0mm
		},
        nodes near coords,
        every node near coord/.append style={
								   font=\footnotesize,
								   /pgf/number format/.cd,
								   	fixed zerofill,
									1000 sep={},
									precision=2
								   },
        enlarge x limits=0.18,
    ]
\addplot[fill=blue,opacity=1.00] 
coordinates {
($m=2$, 0.985)
($m=3$, 0.995)
($m=4$, 0.995)
($m=5$, 0.995)
};
\end{axis}
\end{tikzpicture}
      &
\begin{tikzpicture}[scale=0.6, every node/.style={scale=1.0}]
    \begin{axis}[
        width  = 0.37*\textwidth,
        height = 5.25cm,
        ymin=0,ymax=1.14,
        ytick={0.0,0.2,0.4,0.6,0.8,1.0},
        major x tick style = transparent,
        ybar=5*\pgflinewidth,
        bar width=22pt,
        xlabel = {Number of mixture components},
        ylabel = {AUC},
        xlabel shift = 6pt,
        symbolic x coords={
$m=2$,
$m=3$,
$m=4$,
$m=5$
	},
        xtick={
$m=2$,
$m=3$,
$m=4$,
$m=5$
	},
	y tick label style={
		font=\small,
    		/pgf/number format/.cd,
   		fixed,
   		fixed zerofill,
		1000 sep={},
    		precision=2},
        x tick label style={
		font=\footnotesize,
		inner sep=0mm
		},
        nodes near coords,
        every node near coord/.append style={
								   font=\footnotesize,
								   /pgf/number format/.cd,
								   	fixed zerofill,
									1000 sep={},
									precision=2
								   },
        enlarge x limits=0.18,
    ]
\addplot[fill=blue,opacity=1.00] 
coordinates {
($m=2$, 0.995)
($m=3$, 0.99)
($m=4$, 1.00)
($m=5$, 0.995)
};
\end{axis}
\end{tikzpicture} \\
      \multicolumn{2}{c}{(b) $\mbox{Window size} = 256$} \\ \\[-2ex]
      {\scriptsize\hspace*{0.25in} Test on Zbot}
      & 
      {\scriptsize\hspace*{0.25in} Test on Zeroaccess} \\[-0.25ex]
\begin{tikzpicture}[scale=0.6, every node/.style={scale=1.0}]
    \begin{axis}[
        width  = 0.37*\textwidth,
        height = 5.25cm,
        ymin=0,ymax=1.14,
        ytick={0.0,0.2,0.4,0.6,0.8,1.0},
        major x tick style = transparent,
        ybar=5*\pgflinewidth,
        bar width=22pt,
        xlabel = {Number of mixture components},
        ylabel = {AUC},
        xlabel shift = 6pt,
        symbolic x coords={
$m=2$,
$m=3$,
$m=4$,
$m=5$
	},
        xtick={
$m=2$,
$m=3$,
$m=4$,
$m=5$
	},
	y tick label style={
		font=\small,
    		/pgf/number format/.cd,
   		fixed,
   		fixed zerofill,
		1000 sep={},
    		precision=2},
        x tick label style={
		font=\footnotesize,
		inner sep=0mm
		},
        nodes near coords,
        every node near coord/.append style={
								   font=\footnotesize,
								   /pgf/number format/.cd,
								   	fixed zerofill,
									1000 sep={},
									precision=2
								   },
        enlarge x limits=0.18,
    ]
\addplot[fill=blue,opacity=1.00] 
coordinates {
($m=2$, 0.995)
($m=3$, 1.00)
($m=4$, 1.00)
($m=5$, 1.00)
};
\end{axis}
\end{tikzpicture}
      &
\begin{tikzpicture}[scale=0.6, every node/.style={scale=1.0}]
    \begin{axis}[
        width  = 0.37*\textwidth,
        height = 5.25cm,
        ymin=0,ymax=1.14,
        ytick={0.0,0.2,0.4,0.6,0.8,1.0},
        major x tick style = transparent,
        ybar=5*\pgflinewidth,
        bar width=22pt,
        xlabel = {Number of mixture components},
        ylabel = {AUC},
        xlabel shift = 6pt,
        symbolic x coords={
$m=2$,
$m=3$,
$m=4$,
$m=5$
	},
        xtick={
$m=2$,
$m=3$,
$m=4$,
$m=5$
	},
	y tick label style={
		font=\small,
    		/pgf/number format/.cd,
   		fixed,
   		fixed zerofill,
		1000 sep={},
    		precision=2},
        x tick label style={
		font=\footnotesize,
		inner sep=0mm
		},
        nodes near coords,
        every node near coord/.append style={
								   font=\footnotesize,
								   /pgf/number format/.cd,
								   	fixed zerofill,
									1000 sep={},
									precision=2
								   },
        enlarge x limits=0.18,
    ]
\addplot[fill=blue,opacity=1.00] 
coordinates {
($m=2$, 1.00)
($m=3$, 1.00)
($m=4$, 1.00)
($m=5$, 1.00)
};
\end{axis}
\end{tikzpicture} \\
      \multicolumn{2}{c}{(c) $\mbox{Window size} = 128$} \\ \\[-2ex]
      \end{tabular}
   \caption{Entropy vs window size for Winwebsec models}
   \label{fig:roc-windowsize-winwebsec}
\end{figure}

\begin{figure}[!htb]
      \centering
      \begin{tabular}{cc}
      {\scriptsize\hspace*{0.25in} Test on Zbot}
      & 
      {\scriptsize\hspace*{0.25in} Test on Winwebsec} \\[-0.25ex]
\begin{tikzpicture}[scale=0.6, every node/.style={scale=1.0}]
    \begin{axis}[
        width  = 0.37*\textwidth,
        height = 5.25cm,
        ymin=0,ymax=1.14,
        ytick={0.0,0.2,0.4,0.6,0.8,1.0},
        major x tick style = transparent,
        ybar=5*\pgflinewidth,
        bar width=22pt,
        xlabel = {Number of mixture components},
        ylabel = {AUC},
        xlabel shift = 6pt,
        symbolic x coords={
$m=2$,
$m=3$,
$m=4$,
$m=5$
	},
        xtick={
$m=2$,
$m=3$,
$m=4$,
$m=5$
	},
	y tick label style={
		font=\small,
    		/pgf/number format/.cd,
   		fixed,
   		fixed zerofill,
		1000 sep={},
    		precision=2},
        x tick label style={
		font=\footnotesize,
		inner sep=0mm
		},
        nodes near coords,
        every node near coord/.append style={
								   font=\footnotesize,
								   /pgf/number format/.cd,
								   	fixed zerofill,
									1000 sep={},
									precision=2
								   },
        enlarge x limits=0.18,
    ]
\addplot[fill=blue,opacity=1.00] 
coordinates {
($m=2$, 0.66)
($m=3$, 0.65)
($m=4$, 0.71)
($m=5$, 0.72)
};
\end{axis}
\end{tikzpicture}
      &
\begin{tikzpicture}[scale=0.6, every node/.style={scale=1.0}]
    \begin{axis}[
        width  = 0.37*\textwidth,
        height = 5.25cm,
        ymin=0,ymax=1.14,
        ytick={0.0,0.2,0.4,0.6,0.8,1.0},
        major x tick style = transparent,
        ybar=5*\pgflinewidth,
        bar width=22pt,
        xlabel = {Number of mixture components},
        ylabel = {AUC},
        xlabel shift = 6pt,
        symbolic x coords={
$m=2$,
$m=3$,
$m=4$,
$m=5$
	},
        xtick={
$m=2$,
$m=3$,
$m=4$,
$m=5$
	},
	y tick label style={
		font=\small,
    		/pgf/number format/.cd,
   		fixed,
   		fixed zerofill,
		1000 sep={},
    		precision=2},
        x tick label style={
		font=\footnotesize,
		inner sep=0mm
		},
        nodes near coords,
        every node near coord/.append style={
								   font=\footnotesize,
								   /pgf/number format/.cd,
								   	fixed zerofill,
									1000 sep={},
									precision=2
								   },
        enlarge x limits=0.18,
    ]
\addplot[fill=blue,opacity=1.00] 
coordinates {
($m=2$, 0.99)
($m=3$, 0.99)
($m=4$, 0.92)
($m=5$, 0.82)
};
\end{axis}
\end{tikzpicture} \\
      \multicolumn{2}{c}{(a) $\mbox{Window size} = 512$} \\ \\[-2ex]
      {\scriptsize\hspace*{0.25in} Test on Zbot}
      & 
      {\scriptsize\hspace*{0.25in} Test on Winwebsec} \\[-0.25ex]
\begin{tikzpicture}[scale=0.6, every node/.style={scale=1.0}]
    \begin{axis}[
        width  = 0.37*\textwidth,
        height = 5.25cm,
        ymin=0,ymax=1.14,
        ytick={0.0,0.2,0.4,0.6,0.8,1.0},
        major x tick style = transparent,
        ybar=5*\pgflinewidth,
        bar width=22pt,
        xlabel = {Number of mixture components},
        ylabel = {AUC},
        xlabel shift = 6pt,
        symbolic x coords={
$m=2$,
$m=3$,
$m=4$,
$m=5$
	},
        xtick={
$m=2$,
$m=3$,
$m=4$,
$m=5$
	},
	y tick label style={
		font=\small,
    		/pgf/number format/.cd,
   		fixed,
   		fixed zerofill,
		1000 sep={},
    		precision=2},
        x tick label style={
		font=\footnotesize,
		inner sep=0mm
		},
        nodes near coords,
        every node near coord/.append style={
								   font=\footnotesize,
								   /pgf/number format/.cd,
								   	fixed zerofill,
									1000 sep={},
									precision=2
								   },
        enlarge x limits=0.18,
    ]
\addplot[fill=blue,opacity=1.00] 
coordinates {
($m=2$, 0.638)
($m=3$, 0.536)
($m=4$, 0.52)
($m=5$, 0.69)
};
\end{axis}
\end{tikzpicture}
      &
\begin{tikzpicture}[scale=0.6, every node/.style={scale=1.0}]
    \begin{axis}[
        width  = 0.37*\textwidth,
        height = 5.25cm,
        ymin=0,ymax=1.14,
        ytick={0.0,0.2,0.4,0.6,0.8,1.0},
        major x tick style = transparent,
        ybar=5*\pgflinewidth,
        bar width=22pt,
        xlabel = {Number of mixture components},
        ylabel = {AUC},
        xlabel shift = 6pt,
        symbolic x coords={
$m=2$,
$m=3$,
$m=4$,
$m=5$
	},
        xtick={
$m=2$,
$m=3$,
$m=4$,
$m=5$
	},
	y tick label style={
		font=\small,
    		/pgf/number format/.cd,
   		fixed,
   		fixed zerofill,
		1000 sep={},
    		precision=2},
        x tick label style={
		font=\footnotesize,
		inner sep=0mm
		},
        nodes near coords,
        every node near coord/.append style={
								   font=\footnotesize,
								   /pgf/number format/.cd,
								   	fixed zerofill,
									1000 sep={},
									precision=2
								   },
        enlarge x limits=0.18,
    ]
\addplot[fill=blue,opacity=1.00] 
coordinates {
($m=2$, 0.994)
($m=3$, 0.942)
($m=4$, 0.842)
($m=5$, 0.81)
};
\end{axis}
\end{tikzpicture} \\
      \multicolumn{2}{c}{(b) $\mbox{Window size} = 256$} \\ \\[-2ex]
      {\scriptsize\hspace*{0.25in} Test on Zbot}
      & 
      {\scriptsize\hspace*{0.25in} Test on Winwebsec} \\[-0.25ex]
\begin{tikzpicture}[scale=0.6, every node/.style={scale=1.0}]
    \begin{axis}[
        width  = 0.37*\textwidth,
        height = 5.25cm,
        ymin=0,ymax=1.14,
        ytick={0.0,0.2,0.4,0.6,0.8,1.0},
        major x tick style = transparent,
        ybar=5*\pgflinewidth,
        bar width=22pt,
        xlabel = {Number of mixture components},
        ylabel = {AUC},
        xlabel shift = 6pt,
        symbolic x coords={
$m=2$,
$m=3$,
$m=4$,
$m=5$
	},
        xtick={
$m=2$,
$m=3$,
$m=4$,
$m=5$
	},
	y tick label style={
		font=\small,
    		/pgf/number format/.cd,
   		fixed,
   		fixed zerofill,
		1000 sep={},
    		precision=2},
        x tick label style={
		font=\footnotesize,
		inner sep=0mm
		},
        nodes near coords,
        every node near coord/.append style={
								   font=\footnotesize,
								   /pgf/number format/.cd,
								   	fixed zerofill,
									1000 sep={},
									precision=2
								   },
        enlarge x limits=0.18,
    ]
\addplot[fill=blue,opacity=1.00] 
coordinates {
($m=2$, 0.684)
($m=3$, 0.772)
($m=4$, 0.772)
($m=5$, 0.764)
};
\end{axis}
\end{tikzpicture}
      &
\begin{tikzpicture}[scale=0.6, every node/.style={scale=1.0}]
    \begin{axis}[
        width  = 0.37*\textwidth,
        height = 5.25cm,
        ymin=0,ymax=1.14,
        ytick={0.0,0.2,0.4,0.6,0.8,1.0},
        major x tick style = transparent,
        ybar=5*\pgflinewidth,
        bar width=22pt,
        xlabel = {Number of mixture components},
        ylabel = {AUC},
        xlabel shift = 6pt,
        symbolic x coords={
$m=2$,
$m=3$,
$m=4$,
$m=5$
	},
        xtick={
$m=2$,
$m=3$,
$m=4$,
$m=5$
	},
	y tick label style={
		font=\small,
    		/pgf/number format/.cd,
   		fixed,
   		fixed zerofill,
		1000 sep={},
    		precision=2},
        x tick label style={
		font=\footnotesize,
		inner sep=0mm
		},
        nodes near coords,
        every node near coord/.append style={
								   font=\footnotesize,
								   /pgf/number format/.cd,
								   	fixed zerofill,
									1000 sep={},
									precision=2
								   },
        enlarge x limits=0.18,
    ]
\addplot[fill=blue,opacity=1.00] 
coordinates {
($m=2$, 0.716)
($m=3$, 0.598)
($m=4$, 0.558)
($m=5$, 0.544)
};
\end{axis}
\end{tikzpicture} \\
      \multicolumn{2}{c}{(c) $\mbox{Window size} = 128$} \\ \\[-2ex]
      \end{tabular}
   \caption{Entropy vs window size for Zeroaccess models}
   \label{fig:roc-windowsize-zeroaccess}
\end{figure}

\end{document}